\documentclass[10pt,twocolumn,showpacs,amsmath,amsfonts,pra,superscriptaddress,aps,floatfix]{revtex4-1}

\usepackage[protrusion=true,expansion]{microtype}

\pdfpageheight\paperheight
\pdfpagewidth\paperwidth

\usepackage{xargs}
\usepackage{ifthen}
\usepackage{etoolbox}

\usepackage{mathtools}
\usepackage{mathrsfs}

\usepackage{graphicx}

\usepackage{xspace}

\newcommand*{\ie}{i.e.\@\xspace}
\newcommand*{\eg}{e.g.\@\xspace}
\newcommand*{\cf}{cf.\@\xspace}

\newcommand*{\iid}{i.i.d.\@\xspace}
\makeatletter
\newcommand*{\etc}{%
    \@ifnextchar{.}%
        {etc}%
        {etc.\@\xspace}%
}
\makeatother

\let\originalleft\left
\let\originalmiddle\middle
\let\originalright\right
\renewcommand{\left}{\mathopen{}\mathclose\bgroup\originalleft}
\renewcommand{\middle}{\originalmiddle}
\renewcommand{\right}{\aftergroup\egroup\originalright}

\newcommand{\ii}{\ensuremath{\mathrm{i}}}
\newcommand{\ee}{\ensuremath{\mathrm{e}}}

\DeclareMathOperator{\Tr}{Tr}
\DeclarePairedDelimiter\abs{\lvert}{\rvert}
\DeclarePairedDelimiterX\expval[1]{\langle}{\rangle}{#1}
\DeclarePairedDelimiterX\bra[1]{\langle}{\vert}{#1}
\DeclarePairedDelimiterX\ket[1]{\vert}{\rangle}{#1}
\DeclarePairedDelimiterX\overlap[2]{\langle}{\rangle}{#1\delimsize\vert#2}
\DeclarePairedDelimiterX\bracket[3]{\langle}{\rangle}{#1\delimsize\vert#2\delimsize\vert#3}
\DeclarePairedDelimiterX\clebschgordan[3]{\langle}{\rangle}{#1\,#2\delimsize\vert#3}

\renewcommand{\vec}[1]{\ensuremath{\boldsymbol{#1}}}
\newcommand{\uvec}[1]{\ensuremath{\widehat{\vec{#1}}}}
\newcommand{\ten}[1]{\ensuremath{\boldsymbol{#1}}}
\newcommand{\oten}{\ensuremath{\boldsymbol{1}}}

\newcommand{\Htwo}{\ensuremath{H_{\mathrm{dd}}}}
\newcommand{\mueg}{\ensuremath{\mu_{\mathrm{e} \mathrm{g}}}}

\renewcommandx*\deg[2][1=i,2=,usedefault]{%
  \ifthenelse{\equal{#2}{}}%
    {\ensuremath{\vec{d}_{\mathrm{e} \mathrm{g}}^{#1}}}%
    {\ensuremath{d_{\mathrm{e} \mathrm{g}}^{#1} \left(#2\right)}}%
}
\newcommandx*\dge[2][1=i,2=,usedefault]{%
  \ifthenelse{\equal{#2}{}}%
    {\ensuremath{\vec{d}_{\mathrm{g} \mathrm{e}}^{#1}}}%
    {\ensuremath{d_{\mathrm{g} \mathrm{e}}^{#1} \left(#2\right)}}%
}

\newcommandx*\Pg[1][1=i]{\ensuremath{P_{\mathrm{g}}^{#1}}}
\newcommandx*\Pe[1][1=i]{\ensuremath{P_{\mathrm{e}}^{#1}}}

\renewcommandx*\S[2][1=i,2=j]{\ensuremath{\ten{S}_{#1 #2}}}
\newcommandx*\C[2][1=i,2=j]{\ensuremath{\ten{C}_{#1 #2}}}

\newcommand{\Je}{\ensuremath{J_{\mathrm{e}}}}
\newcommand{\me}{\ensuremath{m_{\mathrm{e}}}}
\newcommand{\Jg}{\ensuremath{J_{\mathrm{g}}}}
\newcommand{\mg}{\ensuremath{m_{\mathrm{g}}}}

\newcommandx*\nS[3][1=n,2=\Jg{},3=\mg{},usedefault=@]{\ensuremath{#1 \mathrm{S}_{#2\ifstrempty{#3}{}{,#3}}}}
\newcommandx*\nP[3][1=n,2=\Je{},3=\me{},usedefault=@]{\ensuremath{#1 \mathrm{P}_{#2\ifstrempty{#3}{}{,#3}}}}

\begin{document}

\title{The Spectral Backbone of Excitation Transport in Ultra-Cold Rydberg Gases}

\author{Torsten Scholak}
\email[Please address correspondence to: ]{torsten.scholak@googlemail.com}
\affiliation{Chemical Physics Theory Group, Department of Chemistry and Center for Quantum Information and Quantum Control, University of Toronto, Toronto, Canada M5S 3H6}
\affiliation{Physikalisches Institut der Albert-Ludwigs-Universit\"at, Hermann-Herder-Str.~3, D-79104 Freiburg, Germany}

\author{Thomas Wellens}
\affiliation{Physikalisches Institut der Albert-Ludwigs-Universit\"at, Hermann-Herder-Str.~3, D-79104 Freiburg, Germany}

\author{Andreas Buchleitner}
\affiliation{Physikalisches Institut der Albert-Ludwigs-Universit\"at, Hermann-Herder-Str.~3, D-79104 Freiburg, Germany}
\affiliation{Freiburg Institute for Advanced Studies, Albert-Ludwigs-Universit\"at, Albertstr.~19, D-79104 Freiburg, Germany}

\date{\today}


\begin{abstract}
  The spectral structure underlying excitonic energy transfer in ultra-cold Rydberg gases is studied numerically, in the framework of random matrix theory, and via self-consistent diagrammatic techniques. Rydberg gases are made up of randomly distributed, highly polarizable atoms that interact via strong dipolar forces. Dynamics in such a system is fundamentally different from cases in which the interactions are of short range, and is ultimately determined by the spectral and eigenvector structure. In the energy levels' spacing statistics, we find evidence for a critical energy that separates delocalized eigenstates from states that are localized at pairs or clusters of atoms separated by less than the typical nearest-neighbor distance. We argue that the dipole blockade effect in Rydberg gases can be leveraged to manipulate this transition across a wide range: As the blockade radius increases, the relative weight of localized states is reduced. At the same time, the spectral statistics---in particular, the density of states and the nearest neighbor level spacing statistics---exhibits a transition from approximately a $1$-stable L\'{e}vy to a Gaussian orthogonal ensemble. Deviations from random matrix statistics are shown to stem from correlations between inter-atomic interaction strengths that lead to an asymmetry of the spectral density and profoundly affect localization properties. We discuss approximations to the self-consistent Matsubara-Toyozawa locator expansion that incorporate these effects.
\end{abstract}

\maketitle

\section{Introduction}

\label{sec:introduction}

When a gas of neutral atoms is excited into weakly bound Rydberg states, exceptionally large interactions due to enhanced polarizability occur. Attractive forces accelerate atoms towards each other and energy is exchanged resonantly in binary collisions---a process studied in great detail during the past forty years \cite{Gallagher:2005ai}. More recently, laser cooling and trapping techniques have permitted to suspend atomic motion to the extent where energy transfer dynamics is dominated not by collisions of atom pairs, but rather by many-body processes \cite{Gallagher:2008aa,Comparat:2010bv,Akulin:1999zp}. When motional degrees of freedom are frozen out, dipolar interactions can cause coherent redistribution of energy, during which an excitation can delocalize over many atoms and great lengths. In this regime, transfer of excitations in the Rydberg gas shows undeniable similarities to energy transport in L\'{e}vy spin glasses \cite{Binder:1986ht,*Edwards:1975hb,*Mezard:1987lp}, between nitrogen-vacancy centers in diamond \cite{Wrachtrup:2006gd,*Hanson:2008rp,*Balasubramanian:2009rw,*Witzel:2012wq,*Dolde:2014ov}, and in certain molecular aggregates such as, \eg{}, lattice-confined polar molecules \cite{Yan:2013yj} and light-harvesting complexes that are employed in photosynthesis \cite{Amerongen:2000fk,*Engel:2007cr,*Cheng:2009ek,*Nalbach:2010zr,*Fleming:2011tg}. 

An advantage of Rydberg atoms and gases is their high degree of experimental controllability \cite{Anderson:1998cg,Singer:2004gg,Singer:2004zi,Tong:2004no,Heidemann:2007lt,Gaetan:2009jl,Saffman:2010ug,Weimer:2010wj,Anderson:2011zn,Walker:2012sr,Gunter:2013aa}. In particular, radiative losses are weak, what allows to study the regime of purely coherent excitation transfer---or to introduce, in a controlled way, various sources of noise giving rise to predominantly incoherent transport mechanisms, in different degrees of freedom \cite{Ates:2012ce,Muller:2012ft,Olmos:2011fm,*Gunter:2013aa}. Rydberg gases therefore constitute an ideal testbed to study the physics of coherent (or incoherent) energy transfer which, due to the combined influence of disorder and coherence, gives rise to an intriguing variety of phenomena ranging from diffusive (where the excitation, if initially localized at a single atom, eventually spreads over the whole cloud) to localized transport (where, even at long times, the excitation remains localized in a certain sub-region of the cloud) \cite{Anderson:1958qw,Kramer:1993cy,Abrahams:2010pl,Robicheaux:2014vr}. While disorder and interaction-induced transport phenomena represent a long-standing and central theme of condensed matter theory \cite{Anderson:1958qw,Kramer:1993cy,Imry:1997jk,Mott:1970fk}, mesoscopic physics \cite{Geisel:1986aa,Ketzmerick:2000aa,Kottos:1999iz,Paul:2007dd,Geiger:2012kn,Zhao:2014fz}, light matter interaction \cite{Bayfield:1989mu,Koch1995289,Casati:1988hr,Arndt:1991ks,Moore:1994aa,Wimberger:2002aa,Jorder:2014pf,Wellens:2008sp} and, more recently, quantum simulation \cite{Billy:2008mh,Schreiber:2012as,Peruzzo:2010ca}, one can argue that cold Rydberg gases offer the specific advantage to address rather subtle issues of quantum transport theory in disordered systems, which hitherto could not be addressed. In our present contribution, we will focus on the tunability of the spectral structure of these experimental objects.

Since the seminal works of \citet{Anderson:1958qw} and \citet{Abrahams:1979aa}, it has been known that a metal-insulator phase transition exists in three-dimensional lattices with random site energies and short-range interactions \cite{Kramer:1993cy,Wolfle:2010aa}: If, above a certain strength of disorder, the couplings to sites at large distances $R$ decay faster than $R^{-3}$, then all states are exponentially localized, and diffusive transport does not occur---the probability to find an excitation at its site of origin is finite for \emph{all} times. Anderson's result, however, does not apply to the case of dipolar interactions as they occur in a Rydberg gas, where all atoms (and not only nearest neighbors) interact with each other by forces falling off like $R^{-3}$. Previous studies relying on approximate methods like the self-consistent theory of localization \cite{Elyutin:1979qe,Elyutin:1980oq,Logan:1984bh,*Logan:1985lq,Logan:1987qf} and a random matrix approach (implying the neglect of correlations between Hamiltonian matrix elements) \cite{Cizeau:1994ml} indicate the absence of any exponentially localized states in the spectrum. Notwithstanding, whilst the system does not exhibit strict Anderson localization, excitation transport may still appear to be spatially localized on the time scale of observation \cite{Logan:1987qf}. Such transient localization can occur due to the existence of algebraically localized (confluence) states \cite{Brezini:1992ib} giving rise to slow sub-diffusive transport where the excitation continues to spread towards more distant atoms with increasing time, but slower than in case of diffusion. Although algebraic localization has been suggested for $R^{-3-\varepsilon}$ with $\varepsilon > 0$, \cf{} Ref.~\onlinecite{Cizeau:1994ml}, the question whether it also occurs for the borderline case $R^{-3}$ is, to our knowledge, still open \cite{Metz:2010uq}.

It is one aim of our work to resolve this question by means of numerically exact simulations of energy transfer in a large frozen Rydberg gas cloud. As a first step, the present article analyzes the {\em spectrum} of single excitations in the cloud. To give a definite answer to the localization problem, also the corresponding eigenstates must be considered, which is the subject of future work based on the results of this article. Note that, due to the recent progress on direct observation techniques of excitation energy transfer \cite{Hofmann:2013jv,Gunter:2013aa}, there is a significant incentive to explicitly study it, both experimentally and theoretically. Nevertheless, the analysis of the Hamiltonian eigenvalue spectrum (while not easily accessible in such experiments) is of interest on its own: as discussed below, the spectrum itself already carries already information on the spatial extent of eigenfunctions and, therefore, qualitatively also the nature of excitation energy transfer. Furthermore, spectral statistics are interesting from a fundamental theoretical point of view, since they can be compared to universal fluctuation properties of random matrix spectra \cite{Mehta:1991ph}. The assessment of the spectral statistics can be regarded as the first milestone to a complete physical picture of any complex system.

Given the computational resources, a feasible and exact approach to the description of large disordered systems is the numerical diagonalization of an exhaustive number of disorder realizations of the Hamiltonian. Based on this method, we present results for the spectral density and for the nearest-neighbor level spacing distribution in different regions of energy. We indeed find evidence for a localization transition that is seeded by rare, yet important occurrences of strongly coupled pairs of nearby Rydberg atoms. Significantly, localization is shown to be controllable: it depends on the strength of the dipole blockade \cite{Tong:2004no,Singer:2004zi,Liebisch:2005hy} that has been proposed as a means to perform quantum gate operations with Rydberg atoms \cite{Jaksch:2000wl}. It can also be used to influence the distribution of pair distances (\eg{}, to tune the minimal distance between Rydberg atoms) \cite{Reetz-Lamour:2007sa} as well as to create collective Rydberg excitations that spread over many atoms \cite{Comparat:2010bv}. In the blockaded regime, the collective ground state of the system has been predicted to transition from a disordered into an ordered, crystalline phase \cite{Weimer:2008aa} which, as shown here, is accompanied by a reduction and eventual elimination of localization.

We subsequently compare the numerically obtained statistics to results from random matrix theory \cite{Mehta:1991ph} and certain self-consistent perturbative techniques \cite{Brezini:1992ib}. Random matrix theory is widely used in studies of complex systems \cite{Stockmann:1999gd,Beenakker:1997cy,*Mirlin:2000mq,Walschaers:2013fc,*Walschaers:2014va}, where it replaces a complicated Hamiltonian model that is not fully known or not solvable. We specifically discuss the applicability of universal Gaussian orthogonal \cite{Mehta:1991ph} and $\alpha$-stable random matrix ensembles \cite{Cizeau:1994ml}, the latter of which have been studied in the field of L\'{e}vy spin glasses \cite{Neri:2010zk,Janzen:2010ez,*Lemeshko:2013km}. Random matrix theory is typically restricted to matrices with independent and identically distributed entries \cite{Goetschy:2013df}, the Gaussian and $\alpha$-stable ensembles being no exception. Spectral signatures of correlations between Hamiltonian matrix elements that are identified in our numerical reference can thus not be reproduced. A possible remedy is provided by diagrammatic perturbation series, since these account for correlations. Their drawback is, however, that they are limited by a quickly increasing complexity of higher order corrections. Yet, we assess and confirm their capability by solving analytical expressions for the spectral density derived from approximations to the self-consistent ensemble-averaged Matsubara-Toyozawa locator expansion \cite{Matsubara:1961pd,Elyutin:1979qe,Mezard:1999cv}.

Correspondingly, this document is organized as follows: Sec.~\ref{sec:theoretical_description} introduces our model of coherent dipolar energy transfer between resonant levels of ultra-cold Rydberg atoms and further discusses its basic properties. In Sec.~\ref{sec:spectrum}, we numerically analyze the spectral statistics, and argue that atomic proximity and disorder-induced localization phenomena are intertwined with each other. Sec.~\ref{sec:theory} interprets our numerical findings by comparison to theoretical approaches. 
We conclude with Sec.~\ref{sec:conclusions}.

\section{The Rydberg Gas Model And Its Basic Properties}
\label{sec:theoretical_description}

\subsection{The Single-Excitation Transport Hamiltonian}
\label{sec:theoretical_description:hamiltonian}

Let us comment on the conditions under which we consider energy transport. It is a well-known fact \cite{Gallagher:2005ai} that atoms in or close to their ground state constitute an inert, inactive background to resonant transfer of energy in an ultra-cold Rydberg gas. Since this study is dedicated solely to the latter, only the fraction $N$ of the atomic vapor that is excited to Rydberg states will be described. In particular, for each Rydberg atom, we consider a reduced level manifold of a lower and an upper Rydberg state, $\mathrm{S} \equiv \nS[@][1/2][1/2]$ and $\mathrm{P} \equiv \nP[@][3/2][3/2]$, with $n$ being the principal quantum number, the letters $\mathrm{S}$ and $\mathrm{P}$ referring to angular momentum, and the half integers in the subset denoting fine structure. These atomic states are assumed to be energetically well isolated from other eigenstates of $J_{Z}$, the $\uvec{Z}$-component of the total angular momentum operator. In App.~\ref{app:validity_2lvl_approx}, we explain how and to what extent that can be achieved. For Rubidium and $n = 46$, the lifetime of the $\mathrm{S}$ and $\mathrm{P}$ Rydberg states is about or less than $0.1 \mathrm{ms}$ \cite{Beterov:2009ig}. According to the ``frozen Rydberg gas'' hypothesis \cite{Mourachko:1998xw,*Anderson:1998cg,Akulin:1999zp}, for temperatures less than $6 \mu \mathrm{K}$, the thermal motion of the atoms' center of mass coordinates is negligible on this timescale.

Usually, the excitation volume has an elongated, nearly one-dimensional shape with a Gaussian density profile $\rho$, but also homogeneous Rydberg clouds with cigar \cite{Ditzhuijzen:2008kk}, saucer, or spherical shape \cite{Mendonca:2012br} are conceivable. In this article, we concentrate on uniform spherical (true 3D) gases with nearest neighbors separated by typically $\left(2 \pi \rho\right)^{- 1 / 3} \simeq 18.5 \mu \mathrm{m}$ (corresponding to $\rho = 2.5 \times 10^{7} \mathrm{cm}^{-3}$). This distance is large compared to the extent of an electronic wave function (given by $\simeq 0.1 \mu \mathrm{m} \sim n^2 a_{0}$, $a_{0}$ Bohr's radius, for $n = 46$), and electron exchange will be neglected.

Only non-radiative dipole forces proportional to
\begin{align}
  \mathcal{R} \left(R_{i j}\right) & = R_{i j}^{-3}
  \label{eq:H1_Radial}
\end{align}
are taken into account,
\begin{align}
  V \left(\vec{R}_{i j}\right) & = \beta\mathcal{A} \bigl(\uvec{R}_{i j}\bigr) \, \mathcal{R} \left(R_{i j}\right),
  \label{eq:V}
\end{align}
where $\beta$ is a constant depending on the absolute value of the dipole moment of the electronic transition $\mathrm{S} \leftrightarrow \mathrm{P}$, and $\vec{R}_{i j} = \vec{R}_{i} - \vec{R}_{j}$ the distance vector connecting two Rydberg atoms $i$, $j$ with fixed positions $\vec{R}_{i}$, $\vec{R}_{j}$. (For ${}^{85}\mathrm{Rb}$ and $n = 46$, $\beta \simeq 2.59 \times 10^{-14} \mathrm{cm}^2$.) In our approximation, the directional anisotropy $\mathcal{A}$ depends only on the projection $\cos \Theta_{i j} = \uvec{R}_{i j} \cdot \uvec{Z}$ of the orientation $\uvec{R}_{i j} = \vec{R}_{i j} / R_{i j}$ onto the laboratory's $\uvec{Z}$ axis (parallel to the electric field),
\begin{align}
  \mathcal{A} \bigl(\uvec{R}_{i j}\bigr) & = \frac{9 \sqrt{3}}{8 \pi} \left(3 \cos^2 \Theta_{i j} - 1\right).
  \label{eq:H1_Angular}
\end{align}
Note that $V$ is invariant under arbitrary translations of the Rydberg cloud, rotations around the $\uvec{Z}$ axis, and reflections by any mirror parallel to the $\uvec{X}$-$\uvec{Y}$ plane.

Long-range forces $\propto R_{i j}^{-1}$ (mediated by exchange of transversally polarized photons) are negligible as long as the largest inter-atomic distance is small compared to the reciprocal of the transition wavenumber $c / \omega_{0} \simeq 1.20 \mathrm{mm}$, where $\omega_{0}$ is approximately $2 \pi \times 39.7 \mathrm{GHz}$ for the $\mathrm{S}$-$\mathrm{P}$ transition considered here. For the density $\rho = 2.5 \times 10^{7} \mathrm{cm}^{-3}$, we estimate that the near-field approximation for the largest distance (corresponding to the diameter of the cloud) is violated if the cloud contains more than $2 \times 10^{4}$ Rydberg atoms \footnote{However, one can argue that the interaction between atoms separated by such a large distance hardly plays any role at all.}. The analysis in this article is mostly numerical and focuses, also due to computational limitations, on clouds with $N = 10^{4}$ or less atoms. However, at times we also refer to much larger clouds, up to the hypothetical limit $N \to \infty$. These results are (semi-) analytical and provided as approximation to realistically sized clouds.

Since, as mentioned above, relaxation processes (\eg{} due to spontaneous decay) can be neglected, the numbers of $\mathrm{P}$ and $\mathrm{S}$ excitations are conserved. In the following, we assume that, at any point in time, exactly one $\mathrm{P}$ excitation is present in the cloud (among $\left(N - 1\right)$ $\mathrm{S}$ excitations). We thus arrive at the following Hamiltonian:
\begin{align}
  H & = \hbar \omega_{0} \openone + \smashoperator{\sum_{i \neq j = 1}^{N}} \, V \left(\vec{R}_{i j}\right) \, \ket{i} \bra{j}.
  \label{eq:H1}
\end{align}
$\ket*{i}$ denotes the state for which the $\mathrm{P}$ excitation is localized at atom No.~$i$, while all other Rydberg atoms reside in the state $\mathrm{S}$. Keep in mind that, due to a lack of order in the cloud, the numbering and labeling of atoms is arbitrary. That is why the atom labeled $i$ in one realization of the ensemble has no special relation to the atom $i$ in another. The basis $\left\{\ket*{i}\right\}$ spans the so-called single-excitation subspace. The diagonal of $H$ provides only a constant energy shift and is henceforth discarded, $\omega_{0} = 0$. Furthermore, the Hamiltonian, Eq.~\eqref{eq:H1} is scale invariant with respect to the dipole moment $\beta$ and the density $\rho$. In the following, we will therefore rescale energies $\Lambda$ and distances $R$ according to
\begin{subequations}
  \begin{align}
    \Lambda & \mapsto \beta \rho \Lambda = \Lambda \times 6.48 \times 10^{-7} \mathrm{cm}^{-1}
    \shortintertext{and}
    R & \mapsto R / \sqrt[3]{\rho} = R \times 34.2 \mu\mathrm{m},
  \end{align}
\end{subequations}
respectively, or, in other words, choose the units of length and energy such that $\rho \equiv \beta \equiv 1$.

\subsection{Short Distances And Pair Localization}
\label{sec:theoretical_description:pairs}

Due to the dipolar coupling, excitations are exchanged the faster, the closer the Rydberg atoms are. This suggests that in a cluster (or ``cavity'' \cite{Goetschy:2013df}), \ie{}, in a dense agglomeration of two or more atoms, transport occurs much faster than in the more dilute parts of the system. As we discuss now, this implies that energy transport to and away from the cluster is slower than in the case where the cluster is replaced by a single atom.

The smallest cluster is just a pair of two atoms, $i$ and $j$. Consider for the moment that the pair $i j$ shares a distance much smaller than its separation from the surrounding neighbor atoms. Then, also depending on the value of $\mathcal{A} \bigl(\uvec{R}_{i j}\bigr)$, the pair's coupling energy $H_{i j}$ may exceed the interaction strengths to all other atoms by orders of magnitude. In that case, the neighbors' presence plays the role of an insignificant perturbation. The pair-localized states $\left(\ket{i} \pm \ket{j}\right) / \sqrt{2}$ are almost exact eigenstates of the Hamiltonian. These states have approximate eigenenergies
\begin{align}
  \Lambda_{i \pm j} & = \pm H_{i j}.
  \label{eq:lambda_pm}
\end{align}
Any excitation starting from either atom is strongly confined. Likewise, excitations from elsewhere are strongly inhibited from reaching the pair. Henceforth we refer to this effect as ``pair localization''.

\subsection{Probability Density of Matrix Elements}
\label{sec:theoretical_description:randomness}

Transport is only notably affected by the pair localization effect if strongly interacting pairs of nearby Rydberg atoms appear sufficiently frequently. To analyze the likelihood of their occurrence, one has to consider that the randomness in the placement of atoms is passed on deterministically to the Hamiltonian \eqref{eq:H1}, the matrix elements $H_{i j} \equiv \bracket{i}{H}{j} = V \left(\vec{R}_{i j}\right)$ of which thus become random variables themselves. Their statistical properties can be derived on the basis of the constraints we impose on the random picking of atomic positions.

For now, let us set $i$ and $j$ fixed with $i \neq j$. Then $\mathcal{A}$ and $\mathcal{R}$ are independent random variables with known and fairly simple distributions. The statistical distribution of $H_{i j}$, in the following referred to as the probability density function $f_{H_{i j}}$, is then obtained as the distribution of the product $\mathcal{A} \mathcal{R}$. This straight-forward yet cumbersome calculation is carried out in App.~\ref{app:probability_distributions}. From the exact expression, Eq.~\eqref{eq:fH1} (in the limit of a vanishing exclusion radius, the minimal allowed distance between two atoms, $r_{\mathrm{b}} \to 0$), the probability for large interactions (\ie{} short distances) can be derived to scale as follows:
\begin{align}
  f_{H_{i j}} \left(h\right) & \sim_{\abs{h} \to \infty} N^{-1} \abs{h}^{-2}.
  \label{eq:fH1_asympt}
\end{align}
It can be shown that, since $H_{ij} \propto R_{i j}^{-3}$, the power law in $h$ with exponent $-2$ results from the fact that the distribution $f_{R_{i j}}\left(r\right)$ of distances scales like $r^{2}$ for $r \to 0$ (according to the volume element in 3D): $f_{H_{i j}}\left(h\right) \sim f_{R_{i j}}\left(r\right) \, \frac{\mathrm{d}r}{\mathrm{d}h} \sim r^2 r^4 \sim h^{-2}$. The same asymptotics is found for other angular dependencies than \eqref{eq:H1_Angular}, \eg{}, a simplified isotropic model with constant $\mathcal{A}$ or a cloud for which the atomic dipole moments are oriented in all directions fully randomly.

The fact that $f_{H_{i j}}$ decays algebraically according to Eq.~\eqref{eq:fH1_asympt} implies that the occurrence of very large Hamiltonian matrix elements is much more likely than in any Gaussian distribution. In the following, specifically in Secs.~\ref{sec:spectrum:density_of_states} and \ref{sec:spectrum:level_spacing_statistics}, we discuss that localized pairs entail distinctive spectral signatures and are crucial in understanding energy transport in Rydberg clouds. In previous work, it has already been acknowledged that spectral and eigenstate statistics of related physical systems (random dipolar interactions in three dimensions) are affected by pairs \cite{Elyutin:1980oq,Cizeau:1994ml,Akulin:1999zp,Walschaers:2013fc}. However, the details and the extent of their influence are hitherto still largely unknown.

\subsection{Correlations}
\label{sec:theoretical_description:correlations}

The random variables $\left\{H_{i j}\right\}_{i > j}$ are not independent. Consider three tuples of atoms; $i j$, $j k$, and $i k$. The corresponding matrix elements feature correlations that are rooted in basic properties of Euclidean space. Specifically, the triangle inequality 
\begin{align}
  \abs{R_{i j} - R_{j k}} \le R_{i \vphantom{j} k} \le R_{i j} + R_{j k}
  \label{eq:triangle_inequality}
\end{align}
applies to all inter-atomic distances, and correlations thus exist between all matrix elements. That is, the joint probability density function of all matrix elements, $f_{H} \equiv f_{\left\{H_{i j}\right\}_{i > j}}$, does not factor into a product of marginal probability density functions, especially not into the product of single variable densities, $\prod_{k > l} f_{H_{k l}}$ (in which case the $H_{i j}$ are completely independent---a frequent assumption in random matrix theory). The comparison of spectra between models incorporating correlations and those omitting them will be an important part of the upcoming discussion.

\subsection{Dipole Blockade And Short-Range Order}
\label{sec:theoretical_description:blockade}

So far, the model allows the atoms to become arbitrarily close, which is expressed in the algebraic scaling of $f_{H_{i j}}$, Eq.~\eqref{eq:fH1_asympt} above. At first glance, there is no problem in treating the Rydberg atoms as point-sized objects, as long as the randomized inter-atomic distances are sufficiently unlikely to fall below the extent of the respective electronic Rydberg wave-functions. However, that must be reconsidered in light of the existence of the dipole blockade effect \cite{Robicheaux:2005nb}.

The dipole blockade results from van der Waals (\ie{} higher order dipolar) coupling between a Rydberg and a neighboring atom in (or close to) its ground state. In principle, it prevents the laser excitation of the latter into a Rydberg state if the laser's spectral width is small relative to the energy shift induced by the van der Waals coupling \cite{Altshuler:1991si,Lukin:2001ay,Reinhard:2007wq}. The reality is more convoluted than that, however \cite{Ates:2007rp}; in an equally likely scenario, the neighbor and the Rydberg atom may have exchanged places or roles. Since these scenarios are physically indistinguishable, the single Rydberg excitation is eventually shared coherently by all involved atoms. As the excitation process is both collective and localized, these blockaded excitations have been referred to has ``Rydberg super-atoms'' \cite{Robicheaux:2005nb}.

For our purposes, it is sufficient to treat a super-atom like a single Rydberg atom, with the exception that it is equipped with an approximately spherical exclusion volume with radius $r_{\mathrm{b}}$ \cite{Comparat:2010bv}, in which no other Rydberg atom is allowed to exist. In App.~\ref{app:limitations_superatoms}, we discuss some of the limitations of the super-atom and blockade sphere pictures. Due to the nature of the van der Waals coupling, the effect is externally tunable \cite{Ryabtsev:2010zm} and distance dependent, being the stronger the closer the atoms are. In a cloud with many Rydberg excitations, it suppresses the relative likelihood of occurrence of short distances between Rydberg atoms \cite{Vogt:2006if,Comparat:2010bv}.

We therefore introduce the smallest allowed distance, $r_{\mathrm{b}}$, as an additional parameter to our model. In other words, we sample the $N$ atomic positions uniformly within a $3$-dimensional sphere of radius $\sqrt[3]{3 N / \left(4 \pi\right)}$ and only keep those configurations in which all inter-atomic distances are larger than the blockade radius $r_{\mathrm{b}}$. This has two major consequences: First, it is well known that energy transport in ordered systems is fundamentally different from transport in disordered ensembles \cite{Kramer:1993cy,Abrahams:2010pl,Brezini:1992ib}. It is therefore of interest to study large blockade radii $r_{\mathrm{b}}$ as these enforce short-range order. We incorporate random close packing  \cite{Torquato:2000qo} of equisized and non-overlapping hard spheres that, on a phenomenological level, reproduces the recently discovered self-organization of Rydberg super-atoms into crystalline structures \cite{Pohl:2010kb}. Second, the abundance of tightly bound pairs of atoms, as discussed in the previous sections, is conditioned on the power law asymptotics of $f_{H_{i j}}$. For non-vanishing $r_{\mathrm{b}}$, this asymptotics is truncated inasmuch as, according to Eq.~(\ref{eq:V}), the interaction $H_{i j}$ is bounded, $-a / \left(3 r_{\mathrm{b}}^3\right) \leq H_{ij} \leq 2 a / \left(3 r_{\mathrm{b}}^3\right)$ with $a = 27 \sqrt{3} / \left(8 \pi\right)$. This enables a better analysis of effects due to pair localization, since their strength becomes controllable by changing the blockade radius \cite{Urban:2009il,*Gaetan:2009jl}.

\section{Numerical Results: The Spectrum}
\label{sec:spectrum}

In this section, we present numerical results for the spectrum of the single-excitation Rydberg gas Hamiltonian introduced above. We consider, first, the spectral density (averaged over a large number of realizations)---which we show to exhibit a universal power law (for $r_{\mathrm{b}} = 0$) in the tails of the spectrum and a characteristic asymmetry in the center (Sec.~\ref{sec:spectrum:density_of_states})---and second, spectral correlations as described by the nearest neighbor level spacing statistics (Sec.~\ref{sec:spectrum:level_spacing_statistics}). The latter is found to exhibit a transition between Poisson and Wigner-Dyson statistics under variation of the energy (Sec.~\ref{sec:spectrum:mobility_edge}). In Sec.~\ref{sec:spectrum:matrix_correlations}, we analyze the influence of correlations between different matrix elements. In particular, we show that these correlations are responsible for the asymmetry in the spectrum and that they shift the transition from Poisson to Wigner-Dyson statistics towards smaller energies. Finally, the impact of the dipole blockade radius $r_{\mathrm{b}}$ is investigated in Sec.~\ref{sec:spectrum:dipole_blockade} and shown to modify mainly the tails of the spectrum.

\subsection{Spectral Density: Asymptotics And Spectral Center}
\label{sec:spectrum:density_of_states}

\begin{figure}
  \raisebox{-14.40819pt}{\includegraphics{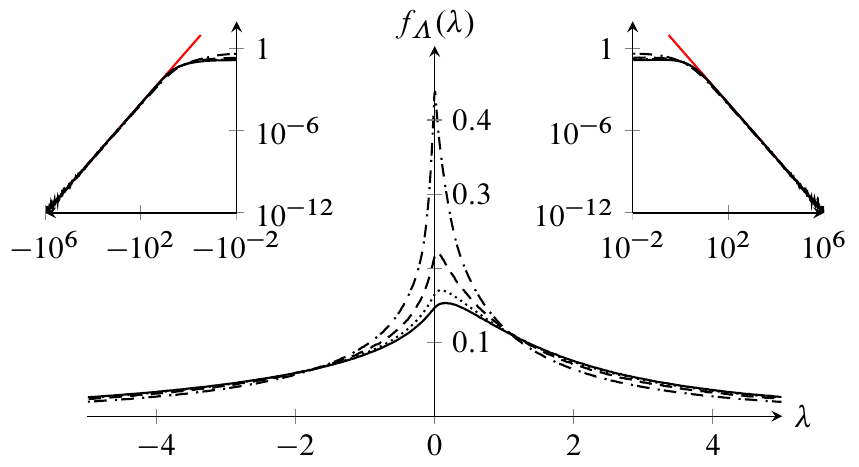}}%
  \caption{(color on-line) Density of states $f_{\Lambda}$ for the ultra-cold Rydberg gases described by the Hamiltonian $H$ in Eq.~\eqref{eq:H1}, with the blockade radius $r_{\mathrm{b}}$ set to zero. The main diagram shows plots for $N = 10$ (dot-dashed; average over $15,660,000$ disorder realizations), $N = 10^{2}$ (dashed; $171,000$ realizations), $N = 10^{3}$ (dotted; $101,550$ realizations), and $N = 10^{4}$ (solid; $22,790$ realizations) Rydberg atoms. The common qualitative features are unimodality (the existence of a single maximum corresponding to the most probable eigenvalue) and a slight skewness to positive energies. The most striking differences are a trend to a smaller central peak and a more pronounced skewness for larger $N$. The log-log plots in the insets illustrate that, asymptotically, $f_{\Lambda} \left(\lambda\right)$ falls off like $\abs{\lambda}^{-2}$ (red/gray) on both sides, irrespective of the system size $N$.}%
  \label{fig:fLambda_rydberg_sizecomp}%
\end{figure}

In this section, we are concerned with the likelihood of certain eigenenergies in our system and how these statistics change with the system's size, $N$. To this end, we study the density of states (spectral density, DOS) that is defined as
\begin{align}
  f_{\Lambda} \left(\lambda\right) & = \frac{1}{N} \, \overline{\Tr \delta \left(\lambda - H\right)}.
\end{align}
$f_{\Lambda} \left(\lambda\right) \, \mathrm{d}\lambda$ is the ensemble averaged probability to find an eigenvalue $\Lambda$ of $H$ in the interval $\left[\lambda, \lambda + \mathrm{d}\lambda\right]$.

Fig.~\ref{fig:fLambda_rydberg_sizecomp} displays histograms obtained by direct numerical diagonalization of many realizations of $H$ in the absence of the dipole blockade effect ($r_{\mathrm{b}} = 0$). These results display the following two characteristic features: 
\begin{itemize}
  \item First, as can be seen in the insets of the figure, the tails of the spectral density obey the same inverse power law for all $N$,
  \begin{align}
    f_{\Lambda} \left(\lambda\right) & \sim_{\abs{\lambda} \to \infty} \abs{\lambda}^{-2}.
    \label{eq:fLambda_asympt}
  \end{align}
  Note, however, that (since the two-level approximation fails for very short atomic distances, \cf{} App.~\ref{app:validity_2lvl_approx}) $f_{\Lambda} \left(\lambda\right)$ for $\abs*{\lambda} \gg 10^{2}$ is, strictly speaking, not physical. The asymptotics of $f_{\Lambda}$ are same as of the probability density $f_{H_{i j}}$ of any matrix element of $H$ (see Eq.~\eqref{eq:fH1_asympt} and the discussion on page \pageref{eq:fH1_asympt} above). This suggests a simple mapping between large matrix elements and extreme eigenvalues, like it was already established in form of Eq.~\eqref{eq:lambda_pm} for the eigenvalues $\Lambda_{i \pm j}$ of pair-localized eigenstates. In other words, only pair-localized states populate the wings of the spectrum. This conclusion is supported by other studies \cite{Elyutin:1980oq,Cizeau:1994ml,Parshin:1998qw,Akulin:1999zp,Neri:2010zk}, in which the close connection between binary interactions with algebraic scaling, rare but extremely strong couplings, extremely large eigenvalues, and heavy-tailed spectral densities has been recognized.
  \item Second, significant differences between the densities of states for different $N$ are found at the spectrum's center, where small systems develop a sharp peak around $\lambda = 0$, and large systems a much more flattened maximum with increased skewness towards positive energies. This skewness and its origin are the subject of the discussion in Sec.~\ref{sec:spectrum:matrix_correlations}. 
\end{itemize}

As noted above, it has been observed in previous studies that Hamiltonians with power-law interactions can have heavy-tailed eigenenergy distributions with infinite variance. In the following, we
\begin{enumerate}
  \renewcommand{\theenumi}{\roman{enumi}}%
  \renewcommand{\labelenumi}{(\theenumi)}%
  \item verify for the Rydberg Hamiltonian \eqref{eq:H1} that this phenomenon corresponds to pairs of eigenstates localized on closely separated, strongly coupled atom pairs (see Sec.~\ref{sec:spectrum:level_spacing_statistics}), and we
  \item provide evidence that the center of the spectrum is occupied by eigenstates that are delocalized over many atoms (also Sec.~\ref{sec:spectrum:level_spacing_statistics}).
\end{enumerate}
Thereafter, we explicitly go beyond the previous studies and
\begin{enumerate}
  \renewcommand{\theenumi}{\roman{enumi}}%
  \renewcommand{\labelenumi}{(\theenumi)}%
  \setcounter{enumi}{2}%
  \item use nearest-neighbor level spacing statistics to find two different transition energies that separate the pair-localized eigenstates in the wings of the spectrum from delocalized states in the spectrum's center (Sec.~\ref{sec:spectrum:mobility_edge}). Furthermore, we
  \item determine that, without the correlations between the Hamiltonian matrix elements, the two transition energies would be much larger and also equal in magnitude (Sec.~\ref{sec:spectrum:matrix_correlations}). Most significantly, we
  \item reveal that, under dipole blockade conditions, the amount of localization is reduced and that the level-spacing statistics undergo a transition upon variation of the dipole blockade radius (Sec.~\ref{sec:spectrum:dipole_blockade}). \label{itm:dipole_blockade_transition}
\end{enumerate}
We regard the last point \eqref{itm:dipole_blockade_transition} the most important new finding of the present work. We expect that the transition of the spectral statistics corresponds to a transition between sub-diffusive and diffuse excitation energy transport, which can be probed with newly introduced excitation population imaging techniques \cite{Gunter:2013aa}, for example. The explicit characterization of energy transport, especially this transition, is the subject of ongoing work.

\subsection{Spectral Correlations: Level Spacing Statistics}
\label{sec:spectrum:level_spacing_statistics}

We have seen that pair-localized states have eigenvalues that are large in absolute value. Evidently, the eigenvalues corresponding to two different pairs are uncorrelated with each other (since the positions of all atoms are drawn independently for each atom). Showing that the edges of the spectrum only contain uncorrelated eigenvalues provides further evidence that all eigenvalues with large absolute value belong to pair-localized states. This is one concern that is addressed in this section. Another is the statistics around the spectrum's center.

Spectral correlations can be analyzed using the nearest-neighbor level spacing density, $f_{S}$, where $f_{S} \left(s\right) \, \mathrm{d}s$ is the ensemble averaged probability to sample two adjacent eigenvalues $\Lambda_{\nu}$, $\Lambda_{\nu + 1}$ that are separated by an energy $S$ between $s$ and $s + \mathrm{d}s$ (for each realization, eigenvalues are sorted in ascending order, $\Lambda_{1} \le \Lambda_{2} \le \ldots \le \Lambda_{N}$). Our data is subject to ``unfolding'' \cite{Dehesa:1984mw} that transforms $f_{S}$ such that the mean level spacing, $\overline{S} = \int_{0}^{\infty} s f_{S} \left(s\right) \, \mathrm{d}s$, is equal to $1$. Thereby, its statistics can be compared to those of the universal Gaussian ensembles that have been extensively studied in the framework of random matrix theory \cite{Wigner:1967nu,Dyson:1970mf,Mehta:1991ph}.

\begin{figure}
  \raisebox{-14.40816pt}{\includegraphics{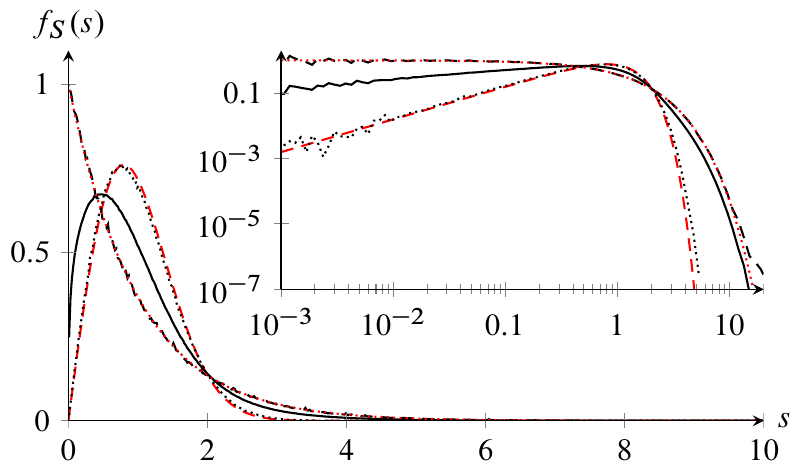}}%
  \caption{(color on-line) Level-spacing density $f_{S}$ for Rydberg gases modeled by Eq.~\eqref{eq:H1} with $N = 10^{4}$ atoms and no dipole blockade, $r_{\mathrm{b}} = 0$. Eigenvalues $\Lambda$ with $\abs{\Lambda} \ge 100$ (dashed) follow Poissonian spacing statistics (red/gray dotted), whereas spacings of eigenvalues from the spectrum's center ($\abs{\Lambda} \le 0.2$, dotted) are distributed according to Rayleigh (Wigner-Dyson) statistics (red/gray dashed). The overall statistics (that is, including all eigenvalues) is mixed and not shown, because it is virtually identical to the spacing statistics in the remaining intervals $\left(- 100, - 0.2\right)$ and $\left(0.2, 100\right)$ (solid). The inset shows a double-logarithmic plot of the same quantities for comparison with algebraic level repulsion for small spacings.}%
  \label{fig:fS_rydberg_sectors}%
\end{figure}

If there were no dependencies between the eigenvalues, all nearest-neighbor distances would be uncorrelated and $f_{S}$ identical to the Poisson distribution,
\begin{align}
  f_{S}^{\mathrm{P}}\left(s\right) & = \ee^{-s}.
  \label{eq:poisson}
\end{align}
Fig.~\ref{fig:fS_rydberg_sectors} shows the spacing statistics for different regions of the spectrum in clouds of $N = 10^{4}$ uniformly distributed, unblockaded (\ie{} $r_{\mathrm{b}}=0$) Rydberg atoms. It is evident that the eigenvalues $\Lambda$ in the wings, $\abs*{\Lambda} \ge 100$, follow Poissonian statistics. This is indeed consistent with pair-localized eigenstates. It can further be assumed that pair localization also occurs away from the wings, since (as can be seen in Fig.~\ref{fig:fS_rydberg_transition} on page \pageref{fig:fS_rydberg_transition}) the level spacing distribution continues to approximate Poissonian statistics for $10 \lesssim \abs*{\Lambda} < 100$.

In App.~\ref{app:validity_2lvl_approx}, we discuss that closely spaced Rydberg atom pairs do not necessarily comply with the two-level approximation that lies at the basis of our calculations. That raises the question of whether or not pair-localization is physical. For the experimental parameters introduced in Sec.~\ref{sec:theoretical_description:hamiltonian}, we have determined that pair-localization is physical for eigenvalues $\Lambda$ with absolute values smaller than roughly $100$. Beyond these energies, it is necessary to consult the eigenenergies (and -states) of the full Hamiltonian. Notwithstanding, there is conclusive evidence that these eigenstates are still localized at the atom pair, albeit not in a superposition of $\ket{\mathrm{S} \mathrm{P}}$ and $\ket{\mathrm{P} \mathrm{S}}$. We arrived at this conclusion by studying the eigenstates of the full three-atom Hamiltonian, with the third atom placed at distances and orientations that are conducive to pair-localization in the case of the Hamiltonian \eqref{eq:H1}.

Let us now continue with the discussion of the level spacing statistics. Towards the center of the spectrum, the spacing statistics switch over smoothly to a Rayleigh distribution,
\begin{align}
  f_{S}^{\mathrm{WD}}\left(s\right) & = \frac{\pi}{2} s \exp \left(-\frac{\pi}{4} s^2\right)
  \label{eq:wigner_dyson}
\end{align}
(see Fig.~\ref{fig:fS_rydberg_transition} and the discussion below), which is also referred to as ``Wigner-Dyson'' statistics in random matrix theory \cite{Wigner:1967nu,Dyson:1970mf}. Let us focus on the frequency of occurrence of small energy differences $s$. A negative deviation from Poissonian statistics signifies ``level repulsion'', which is a manifestation of delocalization \cite{Izrailev:1990qt}. The Wigner-Dyson distribution increases linearly for small spacings, indicating strong level repulsion. 

Let us attempt a short intuitive explanation of why repulsion in the Rydberg level band can be associated with eigenstate delocalization. If two eigenstates occupy approximately the same region in the cloud, then their energies are split. It helps to realize that the physical origin of this split is reminiscent of the avoided level crossing in a two-level atom that is perturbed by an external field. The larger their spatial extent, the more spatial overlap with other states and the more spectral repulsion overall.

\subsection{Critical Level-Spacing Statistics And The Mobility Edge}
\label{sec:spectrum:mobility_edge}

\begin{figure}
  \raisebox{-14.40816pt}{\includegraphics{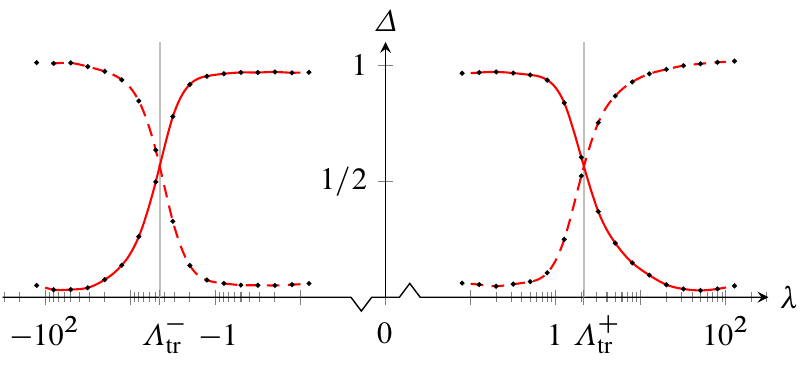}}%
  \caption{(color on-line) Transition between Poissonian and Wigner-Dyson statistics in the level spacing statistics of unblockaded (\ie{} $r_{\mathrm{b}}=0$) Rydberg gases with $N = 10^{4}$ atoms. Shown are the root-mean-square deviations $\Delta\left[f_{S}, f_{S}^{\mathrm{P}}\right]$, $\Delta\left[f_{S}, f_{S}^{\mathrm{WD}}\right]$ to both references, Poisson (red/gray solid, Eq.~\eqref{eq:poisson}) and Wigner (red/gray dashed, Eq.~\eqref{eq:wigner_dyson}), as a function of energy $\lambda$. The dots are numerical data, the curves are polynomial interpolations. The energies of intersection are $\Lambda_{\mathrm{tr}}^{-} = -4.51$ and $\Lambda_{\mathrm{tr}}^{+} = 2.14$.}%
  \label{fig:fS_rydberg_transition}%
\end{figure}

For a quantitative evaluation of the transition between Poissonian and Wigner-Dyson statistics, we consult the root-mean-square (RMS) deviation
\begin{align}
  \Delta \left[f_{S}, g_{S}\right] & = \frac{1}{\xi} \sqrt{\textstyle \int_{0}^{\infty} \, \left[f_{S} \left(s\right) - g_{S} \left(s\right)\right]^{2} \, \mathrm{d}s}
\end{align}
between two spacing densities $f_{S}$ and $g_{S}$. The denominator $\xi$ is such that the deviation $\Delta \left[f_{S}^{\mathrm{P}}, f_{S}^{\mathrm{WD}}\right]$ between the Poisson and the Wigner-Dyson distribution, Eqs.~\eqref{eq:poisson} and \eqref{eq:wigner_dyson}, respectively, is one. Fig.~\ref{fig:fS_rydberg_transition} shows the RMS deviation between the Rydberg level spacing (for $r_{\mathrm{b}} = 0$ and $N = 10^{4}$) and, both, Poissonian and Wigner-Dyson statistics sampled for eigenenergies in logarithmically spaced intervals. The points of intersection between the curves, $\Lambda_{\mathrm{tr}}^{\pm}$, mark the transitions between the statistics. Notably, these are distributed asymmetrically around zero, with $\Lambda_{\mathrm{tr}}^{-} = -4.51$ being bigger in absolute value than $\Lambda_{\mathrm{tr}}^{+} = 2.14$. As discussed above, it is tempting to claim that these transition points are mobility edges separating localized from delocalized states \cite{Mott:1970fk}. A direct verification of this claim requires a detailed analysis of the corresponding eigenstates, which is subject of ongoing work.

\subsection{Correlations between Hamiltonian Matrix Elements}
\label{sec:spectrum:matrix_correlations}

The preceding sections have demonstrated the asymmetry of the spectrum. Asymptotically, the positive and the negative spectrum are identical in absolute value, but the spectrum's mode (\ie{} its most probable eigenvalue) is positive, see Fig.~\ref{fig:fLambda_rydberg_sizecomp}. This is at variance with earlier studies \cite{Parshin:1998qw,Akulin:1999zp} of related models that find the mode to be equal to zero and the spectrum to be fully symmetric around it. As we discuss now, the shortcomings of these studies are related to the suppression or neglect of correlations between the Hamiltonian matrix elements.

\begin{figure}
  \raisebox{-14.40817pt}{\includegraphics{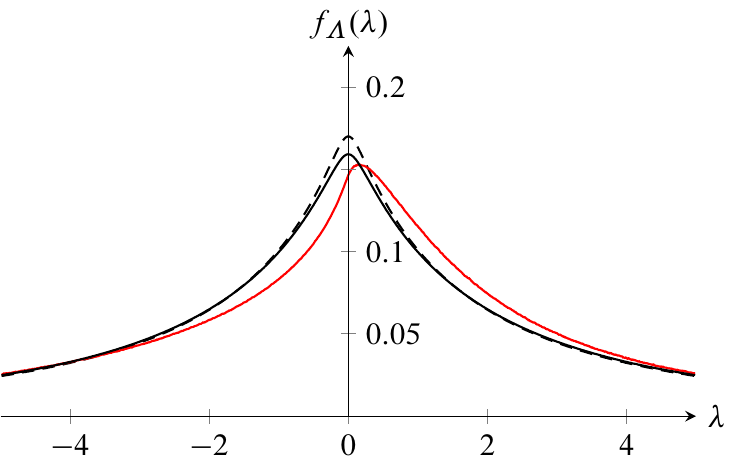}}%
  \caption{(color on-line) Density of states $f_{\Lambda}$ for the Rydberg Hamiltonian \eqref{eq:H1} for $N = 10^{4}$ atoms and $r_{\mathrm{b}} = 0$ (red/gray solid) against the result for the same model but with all correlations set to zero (dashed) and the density of states obtained for the corresponding $1$-stable L\'{e}vy ensemble (black solid), see Sec.~\ref{sec:spectrum:applicability_RMT:SE} below. All densities have the same tail asymptotics, but only the Rydberg model displays the characteristic asymmetry. This shows that the asymmetry originates from correlations between the Rydberg Hamiltonian's matrix elements.}%
  \label{fig:fLambda_stablecomp}%
\end{figure}

We begin by comparing the ensemble-averaged spectral density of the Rydberg gas Hamiltonian \eqref{eq:H1} and a modified Hamiltonian $\tilde{H}$, the matrix elements $\tilde{H}_{i j}$ of which are independently sampled from the marginal probability density $f_{H_{i j}}$ of the original Rydberg Hamiltonian $H$ (with $r_{b} = 0$). The spectral densities that correspond to both models are plotted in Fig.~\ref{fig:fLambda_stablecomp} and agree perfectly when $\abs*{\lambda}$ is large. This illustrates that statistical independence is indeed a valid assumption for the largest elements of $H$ that originate from strongly coupled Rydberg pairs. This is because different pairs are formed in different locations of the cloud.

In the spectrum's center, the modified Rydberg model without correlations turns out to be symmetric  with respect to $\lambda = 0$ (even though $f_{H_{i j}}$ itself is not symmetric, see the inset of Fig.~\ref{fig:probability_distributions} below). This is evidence that the asymmetry of the original Rydberg Hamiltonian is due to the correlations between the matrix elements. We also note that the spectral density of the modified Rydberg model is well reproduced by the $1$-stable (L\'{e}vy) random matrix ensemble (black solid line in Fig.~\ref{fig:fLambda_stablecomp}), which will be introduced in Sec.~\ref{sec:spectrum:applicability_RMT:SE} below.

\begin{table}
  \caption{Transition energies $\Lambda_{\mathrm{tr}}^{\pm}$ for different models describing the exciton dynamics of $N = 10^{4}$ Rydberg atoms. *) Rough estimate, ensemble is not large enough; due to noisy data the statistics deviate from both Poisson and Wigner-Dyson substantially. **) No transition found; statistics closer to Wigner-Dyson at all energies.}%
  \label{tab:transition}%
  \begin{ruledtabular}
	  \begin{tabular}{lllll}
	    model & correlations & $r_{\mathrm{b}}$ & $\Lambda_{\mathrm{tr}}^{-}$ & $\Lambda_{\mathrm{tr}}^{+}$ \\

	    Rydberg & yes & $0$ & $-4.51$ & $2.14$ \\
	    mod.~Rydberg & no & $0$ & $-13.6$ & $13.6$ \\
	    $1$-stable & no & $0$ & $-14.5$ & $14.5$ \\

	    Rydberg & yes & $0.25$ & $-4.84$ & $2.29$ \\
	    mod.~Rydberg & no & $0.25$ & $-14.3$ & $14.3$ \\

	    Rydberg & yes & $0.5$ & $-6.42$ & $3.13$ \\
	    mod.~Rydberg & no & $0.5$ & $-9.45$ *) & $9.46$ *) \\

	    Rydberg & yes & $0.75$ & **) & $3.29$ \\
	    mod.~Rydberg & no & $0.75$ & **) & **) \\
	  \end{tabular}
	\end{ruledtabular}%
\end{table}

Similar conclusions can be drawn from the level spacing statistics, specifically, the transition energies $\Lambda_{\mathrm{tr}}^{\pm}$ that divide the spectrum between Poisson and Wigner-Dyson statistics. Corroborating data is gathered in Tab.~\ref{tab:transition}, where the first three lines correspond to the case $r_{\mathrm{b}} = 0$ without dipole blockade discussed so far. From there it is evident that, for the correlation-free modified Rydberg model, $\Lambda_{\mathrm{tr}}^{-}$ and $\Lambda_{\mathrm{tr}}^{+}$ are equal in absolute value. Furthermore, compared to the correlated case, these transition energies are notably larger in magnitude which means that a greater fraction of the spectrum obeys Wigner-Dyson statistics. In other words, the correlations included in the Rydberg model favor the occurrence of localized states. The physical origin of this effect can be qualitatively understood as follows: imagine a situation where one atom $i$ is much more strongly coupled to another atom $j$ than to all other atoms, $R_{i j} \ll R_{i k} \forall k \neq i,j$. Then, the triangle inequality (\ref{eq:triangle_inequality}) ensures that the same is true for atom $j$, leading to the formation of  a pair-localized state.

\subsection{Dipole Blockade}
\label{sec:spectrum:dipole_blockade}

\begin{figure}
  \raisebox{-230.03804pt}{\includegraphics{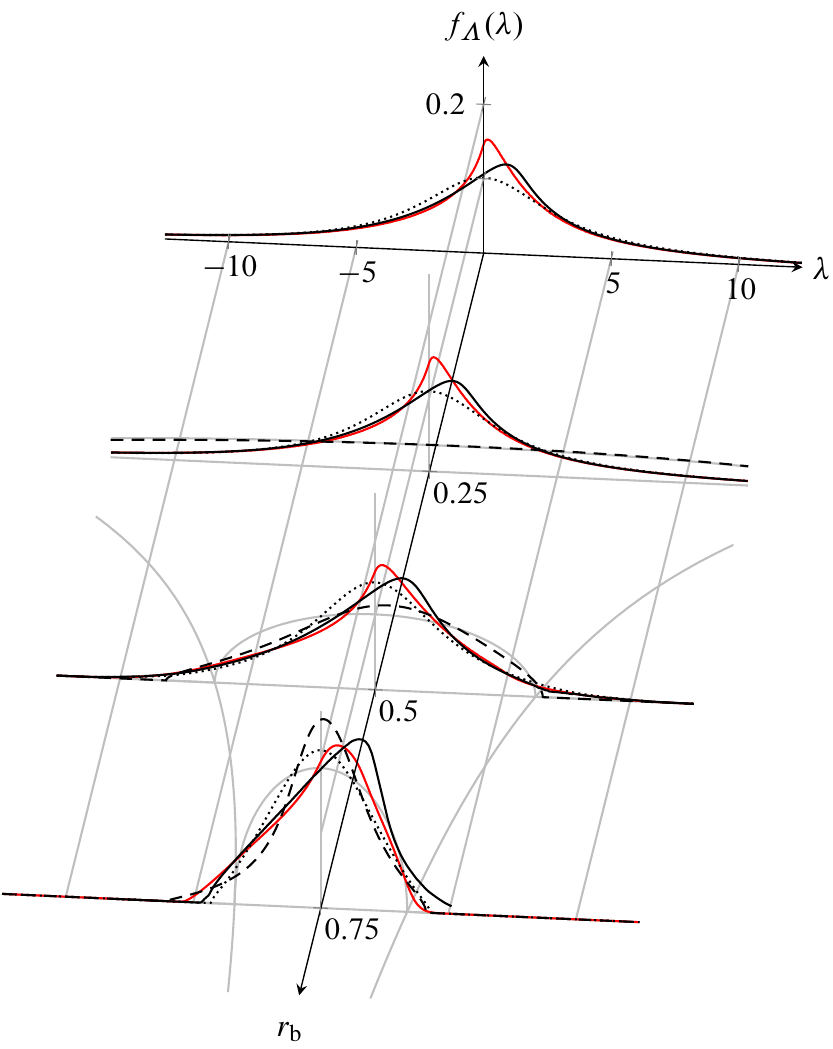}}%
  \caption{(color on-line). Spectral density $f_{\Lambda}$ for $r_{\mathrm{b}} = 0$, $0.25$, $0.5$, and $0.75$. Shown are the respective results of numerically exact diagonalization (red/gray solid) of an ensemble of realizations of Eq.~\eqref{eq:H1} for $N = 10^{4}$ atoms, the low concentration approximation  to the Matsubara-Toyozawa locator expansion  with $l \le 1$ (dotted) as well as $l \le 2$ (black solid), and the approximation to the high concentration locator expansion 
  (dashed). The diagram also depicts the Wigner semicircle law (Eq.~\eqref{eq:wigner_semicircle}) and the boundary of its support (Eq.~\eqref{eq:GOEbounds}) as a function of $r_{\mathrm{b}}$ (both light gray).}%
  \label{fig:fLambda_rydberg_rb_n10000}%
\end{figure}

As mentioned previously, the dipole blockade is expected to have a profound effect on spectral properties, because it inhibits small inter-atomic distances and enforces short-range order. Due to the former, the interaction strengths are capped, their algebraic divergence lifted, and the probability density $f_{H_{i j}}$ has finite support between $-a / \left(3 r_{\mathrm{b}}^3\right)$ and $2 a / \left(3 r_{\mathrm{b}}^3\right)$, where $a = 27 \sqrt{3} / \left(8 \pi\right)$.

In Fig.~\ref{fig:fLambda_rydberg_rb_n10000} (red/gray solid line), we see the development of the spectral density $f_{\Lambda}$ for the Rydberg model \eqref{eq:H1}, when $r_{b}$ is increased from $0$ to $0.75$. (The other curves correspond to theoretical predictions discussed in Sec.~\ref{sec:theory}.) From the graphs it is evident that the inverse power law \eqref{eq:fLambda_asympt} loses its validity in the case of $r_{\mathrm{b}} > 0$; the stronger the blockade, the smaller the width of $f_{\Lambda}$, and the smaller the region in which Eq.~\eqref{eq:fLambda_asympt} applies. In contrast to these significant differences in the wings of the spectrum, the shape of $f_{\Lambda}$ at the spectrum's center is changing only marginally; there is a small increase in skewness and a slight broadening of the central peak.

The nature and strength of the changes suggest that the degree of localization in the system is gravely affected by the raise of the blockade. This is confirmed by studying the level spacing density $f_{S}$, in particular the critical energies $\Lambda_{\mathrm{tr}}^{\pm}$ where the transition between Poisson and Wigner-Dyson statistics takes place. Tab.~\ref{tab:transition} lists these values for $r_{\mathrm{b}} = 0$, $0.25$, $0.5$, and also partially for $0.75$. It can be seen that the transition energies grow slightly in absolute value. It is clear though that these changes are very minor compared to the massive contraction of the wings of the spectrum. For $r_{\mathrm{b}} = 0.75$, the spectrum has been compressed so strongly that a transition between Poisson and Wigner-Dyson statistics on the negative axis does no longer occur. Since the dipole blockade inhibits the short distances needed for pair-localized states, we conjecture that we witness the transition from a partially localized to an almost fully delocalized spectrum.

The data for the modified Rydberg model (also listed in Tab.~\ref{tab:transition}) shows a similar trend and illustrates that the suppression of localized states also happens in the case where the correlations between the matrix elements are all absent. Compared to $r_{\mathrm{b}} = 0$, the $\Lambda_{\mathrm{tr}}^{\pm}$ are somewhat bigger in magnitude for $r_{\mathrm{b}} = 0.25$, just like in the correlated case of model \eqref{eq:H1}. However, this trend is reversed for $r_{\mathrm{b}} = 0.5$ where the interval $(\Lambda_{\mathrm{tr}}^{-}, \Lambda_{\mathrm{tr}}^{+})$ is significantly smaller. This is because it has to fit into the strongly diminished support of the spectrum. For $r_{\mathrm{b}} = 0.75$, the spectrum's support has shrunk once more and a transition cannot be identified; the level spacing statistics are very far away from Poisson and very close to Wigner-Dyson, everywhere in the now comparably narrow spectrum.

We note that these findings are consistent with recent numerical simulations of coherent dipole transport \cite{Robicheaux:2014vr}, where a larger fraction of localized states has been found for increasing degree of randomness in the atomic positions (corresponding to decreasing $r_b$ in our model).

\section{Theoretical Approaches}
\label{sec:theory}

In this section, we compare the numerical results presented in Sec.~\ref{sec:spectrum} with the predictions of various theoretical approaches. As we will see, each of these approaches is able to explain certain features of the numerically obtained spectrum, but none of them suffices to precisely reconstruct the complete spectral density for all values of the blockade radius $r_{\mathrm{b}}$.

\subsection{Random Matrix Theory}
\label{sec:spectrum:applicability_RMT}

In view of the randomness of the Hamiltonian $H$ and the complexity of a large fully interacting system like the Rydberg gas, one should consider whether a statistical ``top-down'' approach is favorable. Such approach would call for random matrix theory (RMT), which is concerned with the statistical properties of eigenvalues and -vectors of large $N \times N$ matrices $M$ with random elements $M_{i j}$ \cite{Mehta:1991ph,Stockmann:1999gd,Akemann:2011aa}. Within RMT, all results are derived from the probability density function $f_{M}$ of $M$. RMT is relevant in theoretical physics, since (i) $M$ can be the matrix representation of the Hamiltonian of a disordered and/or complex system realization and (ii) the statistical properties can, in many cases, be computed analytically. Although, for physical problems, $f_{M}$ is highly nontrivial, commonly impossible to obtain from first principles, and thus simply not known, general, top-down assumptions about the matrix ensemble facilitate tremendous simplifications and produce results with surprisingly universal validity when compared to real experimental data of physical systems such as, \eg{}, complex nuclei \cite{Wigner:1955aa}, chaotic billiards \cite{Stockmann:1999gd}, or strongly perturbed Rydberg systems \cite{Stania:2005aa,Madronero:2005aa}.

\subsubsection{Gaussian Orthogonal Matrices}
\label{sec:spectrum:applicability_RMT:GOE}

\begin{figure}
  \raisebox{-14.40831pt}{\includegraphics{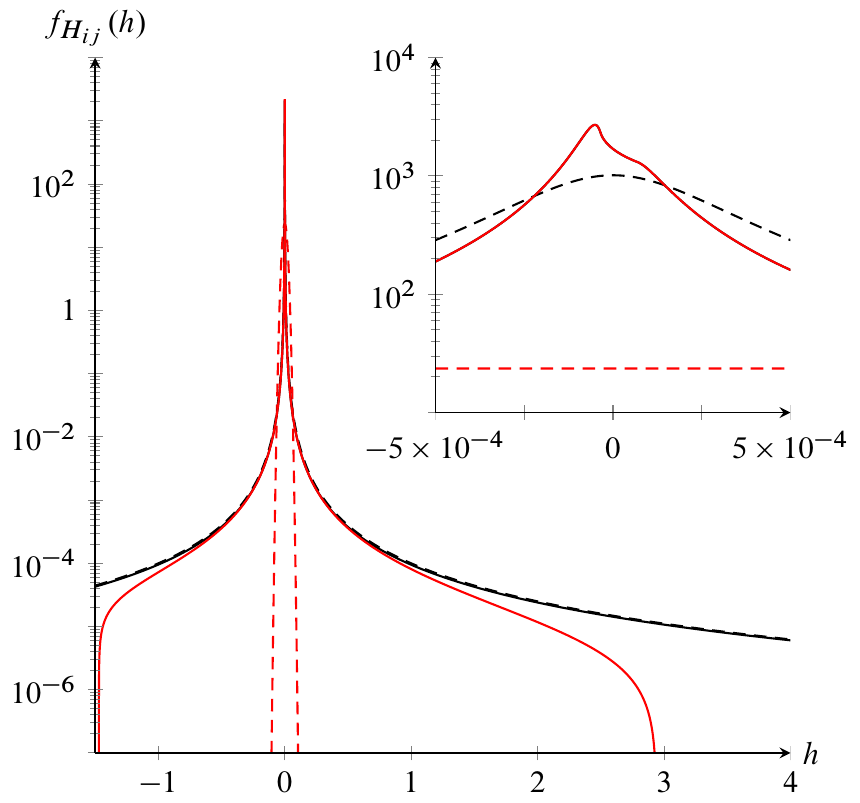}}%
  \caption{(color on-line). Probability density function $f_{H_{i j}}$, see Eq.(\ref{eq:fH1}), for the interactions between $N = 10^{4}$ atoms in the Hamiltonian \eqref{eq:H1}. The black solid curve in the main figure is the graph of Eq.~\eqref{eq:fH1} in the limit $r_{\mathrm{b}} \to 0$. The $1$-stable L\'{e}vy distribution of Eq.~\eqref{eq:fstable} is plotted in black as a dashed line. Both distributions feature the same heavy-tailed asymptotics. In the main figure, they lie on top of each other, but in the inset, which magnifies the small-$\abs*{h}$ region, the graph of $f_{H_{i j}}$ appears lopsided. This is a direct consequence of the anisotropic dipolar interaction, Eq.~\eqref{eq:H1_Angular}. The red/gray solid line is a plot of Eq.~\eqref{eq:fH1} for $r_{\mathrm{b}} = 0.75$. It illustrates the radically different asymptotics for the cases $r_{\mathrm{b}} = 0$ and $r_{\mathrm{b}} > 0$. In contrast, the inset shows that, for small $\abs*{h} < 5 \times 10^{4}$, $f_{H_{i j}}$ is virtually independent of $r_{\mathrm{b}}$---the curve for $r_{\mathrm{b}} = 0.75$ (red/gray solid) covers that for $r_{\mathrm{b}} = 0$ (black solid) completely. For $r_{\mathrm{b}} = 0.75$, the variance of the Hamiltonian matrix elements is finite, $\overline{(H_{i j}){}^2} \simeq 2.86 \times 10^{-4}$. For comparison, a normal distribution with the same variance is plotted as red/gray dashed line.}%
  \label{fig:probability_distributions}%
\end{figure}

One of the simplest and best studied random matrix ensembles is the Gaussian Orthogonal ensemble (GOE), where all elements $M_{i j}$ are real (corresponding to a time-reversal symmetric Hamiltonian) and distributed independently from each other in such a way that the ensemble is invariant under all real orthogonal transformations. Under these assumptions, the matrix elements are \iid{} random variables that follow Gaussian distributions, specifically
\begin{subequations}
  \begin{align}
    f_{M_{i i}} \left(m\right) & = \frac{1}{\sqrt{2 \pi \sigma^2}} \, \exp \left(- \frac{m^2}{2 \sigma^2}\right)
    \intertext{in case of the diagonal and}
    f_{M_{i j}} \left(m\right) & = \frac{1}{\sqrt{\pi \sigma^2}} \, \exp \left(- \frac{m^2}{\sigma^2}\right)\label{eq:GOE_offdiagonal}
  \end{align}
\end{subequations}
in case of the off-diagonal elements. In the limit $N \to \infty$, the density of states $f_{\Lambda}$ of this ensemble is given by the Wigner semicircle law, also known as Wigner's surmise \cite{Wigner:1967nu}. The law reads
\begin{align}
  f_{\Lambda} \left(\lambda\right) & = \begin{dcases}
      \frac{2}{\pi \Lambda_{\mathrm{W}}^2} \sqrt{\Lambda_{\mathrm{W}}^2 - \lambda^2}, & \abs*{\lambda} \le \Lambda_{\mathrm{W}} \\
      0, & \text{else}
    \end{dcases}
    \label{eq:wigner_semicircle}
\end{align}
where $\Lambda_{\mathrm{W}} = \sqrt{2 N} \sigma$. In order to compare Eq.~\eqref{eq:wigner_semicircle} with the spectral density of the Rydberg Hamiltonian, we identify the variance $\sigma^2 / 2$ of Eq.~\eqref{eq:GOE_offdiagonal} with the variance $\overline{(H_{i j}){}^2}$ of the distribution of off-diagonal elements, see Eq.~\eqref{eq:H1_variance} and Fig.~\ref{fig:probability_distributions}, and thereby arrive at
\begin{align}
  \Lambda_{\mathrm{W}} & = \frac{3}{2} \sqrt{\frac{3 b^{6} \left(5 + b^{2} \left(-9 + 4 b\right) + 6 \ln \left(b\right)\right)}{10 N \left(-2 + b^{2} \left(9 - 8 b + b^{4}\right)\right)}}
    \label{eq:GOEbounds}
\end{align}
with $b = \sqrt[3]{6 N / \pi} / r_{\mathrm{b}}$.

For $r_{\mathrm{b}} \to 0$, we obtain $\Lambda_{\mathrm{W}} \to \infty$, and $f_{W} \left(\lambda\right) \to 0$ according to Eq.~\eqref{eq:wigner_semicircle}. Thus, the semicircle law totally fails to reproduce the numerical spectrum for $r_{\mathrm{b}} = 0$. As discussed in Sec.~\ref{sec:theoretical_description:randomness}, this behavior can be traced back to the occurrence of large matrix elements $H_{i j}$ according to the algebraic scaling, Eq.~(\ref{eq:fH1_asympt}). For $r_{\mathrm{b}} > 0$, the probability density $f_{H_{i j}}$ has finite support, see Fig.~\ref{fig:probability_distributions} (red/gray solid line), and thus the variance $\overline{(H_{i j}){}^2}$ is finite. Correspondingly, the numerical spectrum agrees better with the semicircle law for larger $r_{\mathrm{b}}$, see Fig.~\ref{fig:fLambda_rydberg_rb_n10000} for $r_{\mathrm{b}} = 0.5$ and $0.75$ (compare the red solid with the light gray lines). Since, however, the GOE ensemble exhibits no correlations between different matrix elements, it does not reproduce the asymmetry of the spectral density, see Sec.~\ref{sec:spectrum:matrix_correlations}. Furthermore, the GOE ensemble predicts Wigner-Dyson level statistics throughout the entire spectrum \cite{Wigner:1967nu,Dyson:1970mf}, and therefore does not feature a transition between Wigner-Dyson and Poissonian level spacing statistics, see Secs.~\ref{sec:spectrum:level_spacing_statistics} and \ref{sec:spectrum:mobility_edge}. This is due to the fact that the occurrence of very large matrix elements (leading to the formation of pair-localized states) is unlikely according to the Gaussian distribution, Eq.~(\ref{eq:GOE_offdiagonal}).

\subsubsection{Stable Random Matrices}
\label{sec:spectrum:applicability_RMT:SE}

As discussed above, the reason for the failure of the GOE ensemble in the case $r_{\mathrm{b}} = 0$ is the occurrence of large matrix elements in the Rydberg Hamiltonian according to the algebraic scaling law, Eq.~\eqref{eq:fH1_asympt}. In contrast, precisely this behavior is accounted for in  stable random matrix theory (SRMT) \cite{Cizeau:1994ml,Burda:2007mk,Burda:2011aa,Neri:2010zk}. SRMT describes the spectral properties of (infinitely) large symmetric matrices $M$ with L\'{e}vy $\alpha$-stable distributed entries \cite{Nolan:2015yg}, with the distinguished property that the distribution of any linear combination of matrix entries is again $\alpha$-stable. Within the families of stable matrices, the parameter $\alpha$ (called index of stability or characteristic exponent) characterizes the statistical asymptotics of the matrix elements. For $0 < \alpha < 2$, one has $f_{M_{i j}} \left(m\right) \sim_{\abs{m} \to \infty} N^{-1} \abs{m}^{-(1 + \alpha)}$. For $\alpha = 2$, $f_{M_{i j}}$ reduces to the normal distribution, which is why SRMT includes the Gaussian ensemble of RMT as its limit case (a Gaussian distribution is also stable).

For $\alpha = 1$, stable random matrices have the same asymptotics as $H$, \cf{} Ref.~\onlinecite{Cizeau:1994ml}. More precisely, all elements $M_{ij}$ of a $1$-stable matrix $M$ are distributed independently and identically according to
\begin{align}
  f_{M_{i j}} \left(m\right) = \frac{1}{N} \left(m^{2} + \frac{\pi^{2}}{N^{2}}\right)^{-1}.
  \label{eq:fstable}
\end{align}
This probability density function is a Cauchy distribution and describes the statistics of the average of infinitely many \iid{} random variables each with mean $0$ and tail asymptotics as $f_{H_{i j}}$ \cite{Gnedenko:1954yq}. As already mentioned, its asymptotic behavior is identical to the one of the Rydberg Hamiltonian (for $r_{\mathrm{b}} = 0$), whereas it slightly differs from the latter in the center, see Fig.~\ref{fig:probability_distributions}.

For this reason (and the fact that $N$ is large), the $1$-stable matrix ensemble ($1$SE) is very similar to the modified Rydberg Hamiltonian ensemble where, as introduced in Sec.~\ref{sec:spectrum:matrix_correlations}, all correlations between Rydberg Hamiltonian matrix elements are removed. Indeed, both, the spectral density (see Fig.~\ref{fig:fLambda_stablecomp}) and the level spacing statistics (see Tab.~\ref{tab:transition}) coincide well with the results of the modified Rydberg ensemble. In particular, features introduced by pairs (specifically pair localization) are correctly captured by SRMT which, too, fails to reproduce the asymmetry of the spectrum.

Furthermore, it follows from the generalized central limit theorem \cite{Gnedenko:1954yq} that the modified Rydberg gas ensemble lies in the domain of attraction of the $1$SE. In other words, the statistics of both ensembles approach each other further with growing system size $N$ and the spectral densities $f_{\Lambda}$ of both ensembles converge to the limit density derived in Refs.~\onlinecite{Cizeau:1994ml} and \onlinecite{Burda:2007mk}.

The degree of agreement between the Rydberg model \eqref{eq:H1} and the $1$SE in the case $r_{\mathrm{b}} \to 0$ is similar to that of the GOE for larger values of $r_{\mathrm{b}}$. Intermediate $r_{\mathrm{b}}$'s (\eg{} $r_{\mathrm{b}} = 0.25$) are not sufficiently covered by either theory. While tempting, it is not possible to interpolate between the $1$-stable and universal Gaussian RMT by introducing a high-energy cutoff to the density \eqref{eq:fstable}, since the statistics of (infinitely) large matrices with a truncated L\'{e}vy distribution always lie in the basin of attraction of the GOE and therefore make exactly the same predictions as universal Gaussian RMT. However, it might be possible to treat the intermediate case within the general theory of Euclidean random matrices (ERMT) \cite{Mezard:1999cv,Ciliberti:2005kz,Goetschy:2013df,Farhi:1998vs,Amir:2010lc,Skipetrov:2011xb,Krich:2011pz,Mulken:2011yw} that addresses all those very large $N \times N$ matrices $M$ the elements $M_{i j}$ of which depend on pairs $\vec{R}_{i}$, $\vec{R}_{j}$ of $N$ randomly chosen coordinates---precisely as for the Rydberg Hamiltonian, Eq.~\eqref{eq:H1}. Note though that neither ERMT nor any of the other random matrix theories presented so far takes into account correlations between matrix elements \cite{Goetschy:2013df}, the importance of which was highlighted in Sec.~\ref{sec:spectrum:matrix_correlations}. To our knowledge there exists no general, exact method for analytically calculating the spectral density of random matrices with non-\iid{} matrix elements. For this reason, we propose another approach to close the gap that relies on diagrammatic techniques related to the approximate methods typically employed in ERMT.

\subsection{The Locator Expansion}
\label{sec:spectrum:locator_expansion}

The ensemble-averaged spectral density $f_{\Lambda}$ can be expressed in terms of the averaged diagonal elements of the resolvent operator $G \left(z\right) = \left(z - H\right)^{-1}$, $z \in \mathbb{C}$:
\begin{align}
  f_{\Lambda} \left(\lambda\right) = - \frac{1}{\pi} \lim_{\varepsilon \to 0^{+}}
    \Im \, \overline{G_{0 0} \left(\lambda + \ii \varepsilon\right)},
\end{align}
where $\Im$ denotes the imaginary part. In the following, we refer to the first diagonal element of the averaged resolvent only, since all diagonal elements of $G$ obey identical statistics. The analytical approach of this section is based on a self-consistent perturbative expansion of $z \, \overline{G_{0 0} \left(z\right)}$ in powers of $H$ that is called the Matsubara-Toyozawa locator expansion \cite{Matsubara:1961pd,Ziman:1979cs} (not to be confused with Feenberg's self-avoiding walk \cite{Feenberg:1948xv,*Feshbach:1948we,Anderson:1958qw,Abou-Chacra:1973ul}). The exact, infinitely long locator expansion reads
\begin{multline}
  z \, \overline{G_{0 0} \left(z\right)} = 1 + \sum_{l = 1}^{\infty} \sum_{k = l}^{\infty} \frac{\rho^{l}}{z^{k + 1}} \idotsint_{D_{l}} \sum_{\vec{i} \in I_{k l}} \\
    H_{0 i_{1}} H_{i_{1} i_{2}} \dotsm H_{i_{k} 0} \; \mathrm{d} \vec{R}_{1} \dotsm \mathrm{d} \vec{R}_{l}.
  \label{eq:locator_expansion}
\end{multline}
The first summation index $l$ specifies the number of locators, that is, the number of distinct atoms summed over, not including the home atom $0$. $l$ counts to $\infty$, because we consider an infinitely large cloud here (with constant density $\rho\equiv 1$). The domain of integration $D_{l}$ is a $3 l$-dimensional subset of $\mathbb{R}^{3 l}$, where all $l$ integration variables $\vec{R}_{i}$---the $3$-dimensional positions of the atoms $1$ to $l$ that are averaged over---have to fulfill $R_{i} > r_{\mathrm{b}}$ for all $1 \le i \le l$ and $\abs*{\vec{R}_{i} - \vec{R}_{j}} > r_{\mathrm{b}}$ with $1 \le j < i \le l$. The first atom $0$ is always located at the origin. $k + 1$ is equal to the number of interactions $H_{i j}$ in the summand and is always larger than $l$. The symbol $\vec{i} = \left(i_{1}, \ldots, i_{k}\right)$ denotes a multi-index that runs over the $k$-dimensional index set $I_{k l}$, a subset of $\left\{0, 1, \dotsc, l\right\}^{k}$. The following restrictions to this summation apply: (i) Each element of $\vec{i}$, \ie{} the indices $i_{1}$ to $i_{k}$, can be any natural number from $1$ to $l$. (ii) The indices $i_{2}$ to $i_{k - 1}$ can also be equal to $0$. (iii) Successive indices must never be the same, $i_{2} \neq i_{1}, \dotsc, i_{k} \neq i_{k-1}$. (iv) Every integer between $0$ and $l$ has to appear in the multi-index vector $\vec{i}$. For this rule, we may write $\left\{0\right\} \cup \left\{i_{1}\right\} \cup \dotsm \cup \left\{i_{k}\right\} = \left\{0, 1, \dotsc, l\right\}$. (v) The last (and most restrictive) rule says that a vector $\vec{i}$ that remains after applying rules (i-iv) is to be eliminated from the set $I_{k l}$ if it is identical to another one in $I_{k l}$ after a permutation of the alphabet. For instance, $(0,2,1,0)$ is equal to $(0,1,2,0)$ when the alphabet is permuted such that $1$ becomes $2$ and $2$ becomes $1$. Therefore $(0,2,1,0)$ is not summed over. This rule applies because the atomic designation labels $1$ and $2$ are arbitrary after ensemble-averaging.

\begin{figure}
  \includegraphics{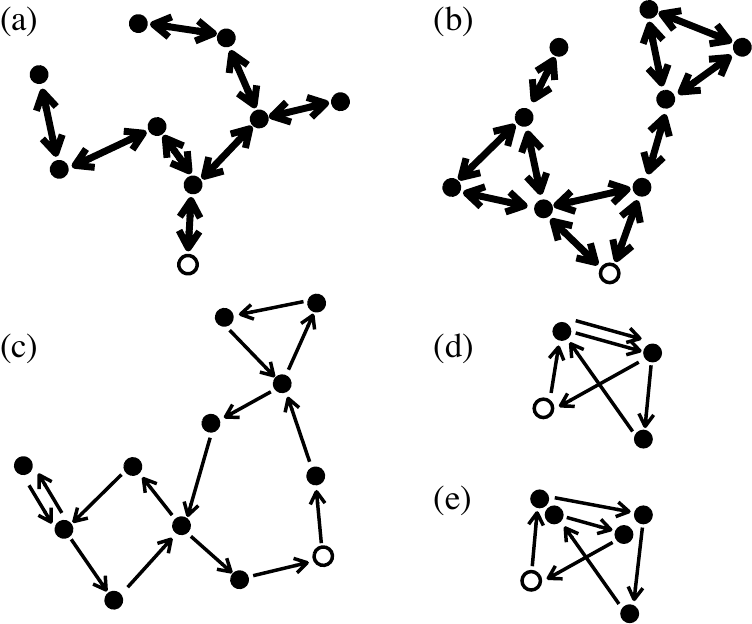}  
  \caption{Examples of paths starting and ending at the home atom (white circle), and thereby contributing to the spectral density. Thin arrows represent single directed transitions from one atom (black or white circle) to another one. A thick double-headed arrow represents arbitrarily many round trip transitions between the two respective atoms. (a) The low concentration limit ($l = 1$) takes into account sequences of arbitrarily many transitions (double-headed thick arrow) between pairs of atoms arranged in a tree-like structure. (b) Low concentration limit ($l = 2$): similar as (a), but involving transitions between three atoms in addition to pairs of atoms. (c) The high concentration limit includes round trips on arbitrarily many atoms each of which is visited only a single time within this round trip. To each atom, however, other round trips may be attached. (d) Example of a path ($0 \to 1 \to 2 \to 3 \to 1 \to 2\to 0$) not taken into account in the classes of paths represented by (a), (b), or (c). (e) Irreducible loop diagram similar to (d).}%
  \label{fig:paths}%
\end{figure}

\citet{Matsubara:1961pd} propose a representation of Eq.~\eqref{eq:locator_expansion} by means of graphs that make it somewhat easier to think and talk about the expansion. In this picture, Eq.~\eqref{eq:locator_expansion} is nothing but a sum of all possible paths or journeys that start and end at the home atom $0$, see Fig.~\ref{fig:paths}. The index $l$ is then the number of distinct intermediate atoms that are visited on the trip. These visits are joined by transition matrix elements $H_{i j}$. For each increase of $l$, different and more complex geometric shapes and diagrams appear. For $l = 1$, we have all repeated back and forth loops between the home and another atom. For $l = 2$, we get all possible transitions between three centers, either arranged in a triangle or spread on a line. And so on. Two approximations to the locator expansion have been derived, the so-called low and high concentration expansions. Both are self-consistent partial summations of Eq.~\eqref{eq:locator_expansion} to infinite order in $H$ over a respective class of diagrams.

\subsubsection{Low Concentration Limit}

\label{sec:spectrum:locator_expansion:low_concentration}

First, recognize that the sum \eqref{eq:locator_expansion} can be renormalized and brought into the self-consistent form
\begin{align}
  z \, \overline{G_{0 0} \left(z\right)} & = 1 + \sum_{l = 1}^{\infty} \rho^{l} F_{l} \bigl(\overline{G_{0 0} \left(z\right)}\bigr)
  \label{eq:locator_expansion:selfconsistent_low}
\end{align}
with $F_{l}$ the ensemble averaged generating function of strongly irreducible graphs with exactly $l + 1$ atoms, respectively \cite{Elyutin:1981pt}. A diagram is called irreducible if it cannot be separated into two independent ones joined by the propagator $1/z$. The irreducible journeys can be thought of as the minimal building blocks of all possible journeys. Eq.~\eqref{eq:locator_expansion:selfconsistent_low} is still exact; the low concentration approximation is obtained by truncating the sum. The name of this approximation is motivated by the prefactor $\rho^{l}$ in Eq.~(\ref{eq:locator_expansion:selfconsistent_low}) which might suggest that higher orders become less important for smaller $\rho$. This conclusion, however, is not always valid since, in general, also $F_l$ depends on $\rho$. This is especially the case for our Rydberg Hamiltonian which is scale-invariant with respect to $\rho$ such that, as already mentioned above, we may set $\rho \equiv 1$.

The first order self-consistent low concentration approximation,
\begin{align}
  z \, \overline{G_{0 0} \left(z\right)} & \simeq 1 + \rho F_{1} \bigl(\overline{G_{0 0} \left(z\right)}\bigr),
  \label{eq:locator_expansion:selfconsistent_low:1st_order}
\end{align}
that includes only the two-center generating function $F_{1}$ generates all graphs that inherit a Cayley tree-like topology with variable connectivity as indicated in Fig.~\ref{fig:paths}(a): they are highly irregular trees with coordination numbers up to $N$ and have connections to neighbors established by an arbitrary high, but equal number of forward and backward links. Loops involving three or more atoms---and hence also correlations between matrix elements due to Eq.~(\ref{eq:triangle_inequality})---are herein neglected.

\begin{figure}
  \raisebox{-14.40817pt}{\includegraphics{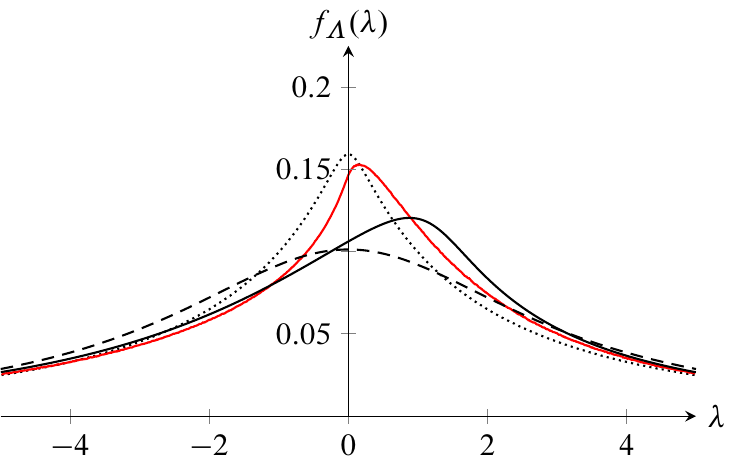}}%
  \caption{(color on-line) Comparison between the numerically obtained density of states $f_{\Lambda}$ for the Hamiltonian \eqref{eq:H1} with parameters $N = 10^{4}$, $r_{\mathrm{b}} = 0$ (red/gray solid line) and the spectral densities derived in Ref.~\onlinecite{Elyutin:1980oq} from the self-consistent Matsubara-Toyozawa locator expansion for low concentrations in first (dashed, Eq.~\eqref{eq:locator_expansion:selfconsistent_low:1st_order}) and second order (black solid, Eq.~\eqref{eq:locator_expansion:selfconsistent_low:2nd_order}) approximation. The dotted line is a plot of the density of states of the $1$-stable L\'{e}vy ensemble.}%
  \label{fig:fLambda_locatorcomp}%
\end{figure}

Let us now solve Eq.~\eqref{eq:locator_expansion:selfconsistent_low:1st_order} for the Hamiltonian \eqref{eq:H1}. The first order generating function on the right side of the equation reads 
\begin{align}
  F_{1} \left(G\right) & = \int_{D_1} P_{1} \left(A_{1}\right) \, \mathrm{d} \vec{R}_{1} .
  \label{eq:F1}
\end{align}
The first order kernel $P_{1}$ represents a sum over arbitrarily many transitions from one atom to another one (at distance $\vec{R}_1$) and back again, \ie{}, $P_{1} \left(A_{1}\right) = \sum_{k = 1}^{\infty} A_{1}^{k} = A_{1} / \left(1 - A_{1}\right)$, with the argument $A_{1} = G^2 \, V^{2} \left(\vec{R}_{1}\right)$, and the Rydberg interaction potential $V$ from Eq.~\eqref{eq:V}. Eq.~(\ref{eq:F1}) thus fully describes all effects (such as pair localization) originating from pairs of atoms. The case $r_{\mathrm{b}} = 0$ is dealt with in Ref.~\onlinecite{Elyutin:1980oq}. In this case, Eq.~\eqref{eq:F1} reduces to $F_{1} \left(G\right) = - \ii \pi G$ for $\Im{G} < 0$. The solution to Eq.~\eqref{eq:locator_expansion:selfconsistent_low:1st_order} then reads  $\overline{G_{0 0} \left(\lambda + \ii \varepsilon\right)} = \left(\lambda + \ii \left(\varepsilon + \pi\right)\right)^{-1}$, so that the predicted spectral density for an infinitely large system is a simple Cauchy distribution with scale $\pi$,
\begin{align}
  f_{\Lambda} \left(\lambda\right) & = \frac{1}{\lambda^{2} + \pi^{2}}.
  \label{eq:locator_expansion:selfconsistent_low:1st_order:solution}
\end{align}
The asymptotics of the tails of this distribution are perfectly correct as can be seen by comparison with Eq.~\eqref{eq:fLambda_asympt} and in Fig.~\ref{fig:fLambda_locatorcomp}. However, the graph of Eq.~\eqref{eq:locator_expansion:selfconsistent_low:1st_order:solution} is less strongly peaked than the numerical reference and lacks the characteristic skewness. The latter is to be expected due to the omission of many-center correlations.

It is interesting to note that the diagrams produced by Eq.~\eqref{eq:locator_expansion:selfconsistent_low:1st_order} are topologically equivalent \cite{Elyutin:1979qe} to those generated by the first order self-consistent approximation of Feenberg's self-avoiding walk \cite{Feenberg:1948xv}. This competing diagrammatic expansion of $\overline{G_{0 0} \left(z\right)}$ is the foundation of the self-consistent theory of localization of \citet{Abou-Chacra:1973ul} and is furthermore leveraged \cite{Cizeau:1994ml,Burda:2011aa} to derive the density of states of the $\alpha$SE in the limit $N \to \infty$. However, despite this deep connection, it is evident from Fig.~\ref{fig:fLambda_locatorcomp} that the spectral density of the $1$SE matches the numerically exact spectral density better than the density \eqref{eq:locator_expansion:selfconsistent_low:1st_order:solution} derived here from the first order low concentration locator expansion \eqref{eq:locator_expansion:selfconsistent_low:1st_order}: While both densities have the correct tail asymptotics, the density of the $1$SE is the better match in the center of the spectrum.

Let us now discuss the case of non-vanishing $r_{\mathrm{b}}$, for which, as discussed at the end of Sec.~\ref{sec:spectrum:applicability_RMT:SE}, $1$-stable random matrix theory is wrong. Direct integration of Eq.~\eqref{eq:F1} leads to
\begin{multline}
  F_{1} \left(G\right) = \frac{8 \pi r_{\mathrm{b}}^{3}}{9} - 3 \sqrt{3} G \sum_{\pm} \left(\frac{r_{\mathrm{b}}^{3}}{2 a G} \mp \frac{1}{3}\right) \\
    \times \sqrt{\frac{1}{3} \pm \frac{r_{\mathrm{b}}^{3}}{a G}} \; \operatorname{arcoth} \sqrt{\frac{1}{3} \pm \frac{r_{\mathrm{b}}^{3}}{a G}}
  \label{eq:F1_solution_blockaded}
\end{multline}
with $a = 27 \sqrt{3} / 8 \pi$. This result makes it impossible to solve Eq.~\eqref{eq:locator_expansion:selfconsistent_low:1st_order} for $G$ analytically. Instead, we have to find the solution numerically for every $z = \lambda + \ii \varepsilon$. We use the covariance matrix adaptation evolution strategy (CMA-ES) \cite{Hansen:2001fu} to find zeros of Eq.~\eqref{eq:locator_expansion:selfconsistent_low:1st_order} with $F_{1}$ as above. We chose $\varepsilon > 0$ large enough such that the algorithm converges, but small enough so that the solution does not depend on $\varepsilon$'s value. The results are plotted in Fig.~\ref{fig:fLambda_rydberg_rb_n10000} (dotted line). Barring the absence of skewness, the method performs well and delivers a decent approximation  to the numerically exact result (red/gray solid line) for model \eqref{eq:H1} for all $r_{\mathrm{b}} > 0$ considered. This is true despite the fact that (i) the locator expansion describes an infinitely large sample, whereas the direct numerical diagonalization is done for the case $N = 10^{4}$, and (ii) that the selection of diagrams constituting the low concentration approximation is primarily geared towards the description of pair localization: With growing minimum inter-atomic distance $r_{\mathrm{b}}$, isolated clusters of few strongly interacting Rydberg atoms become increasingly rare. Instead, each atom interacts with a large variety of other atoms, since the interaction strengths are much more balanced and smaller in absolute value. Therefore, there is no obvious reason why the low concentration approximation should describe the spectral statistics for large $r_{\mathrm{b}}$.

Let us extend the approximation so that it accounts for some of the correlations in $H$. To second order, we obtain
\begin{align}
  z \, \overline{G_{0 0} \left(z\right)} & \simeq 1 + \rho F_{1} \bigl(\overline{G_{0 0} \left(z\right)}\bigr) + \rho^2 F_{2} \bigl(\overline{G_{0 0} \left(z\right)}\bigr).
  \label{eq:locator_expansion:selfconsistent_low:2nd_order}
\end{align}
This self-consistent equation covers all the diagrams of the first order approximation Eq.~\eqref{eq:locator_expansion:selfconsistent_low:1st_order} and beyond that all irreducible three-center loops, generated by the second order generator $F_{2}$, see Fig.~\ref{fig:paths}(b). By invoking combinatorics, Ref.~\onlinecite{Elyutin:1979qe} calculates $F_{2}$ to read
\begin{align}
  F_{2} \left(G\right) & = \iint_{D_{2}} P_{2} \left(A_{1}, A_{1 2}, A_{2}, A_{1 2 3}\right) \, \mathrm{d} \vec{R}_{1} \mathrm{d} \vec{R}_{2}
  \label{eq:F2}
\end{align}
with the kernel $P_{2}$ being equal to
\begin{multline}
  \frac{1}{2} \biggl[\frac{A_{1} + A_{2} + 2 A_{1 2 3}}{1 - \left(A_{1} + A_{1 2} + A_{2} + 2 A_{1 2 3}\right)} \\
  - P_{1} \left(A_{1}\right) - \frac{P_{1} \left(A_{1}\right)}{1 - A_{1}} \left(P_{1} \left(A_{1 2}\right) + P_{1} \left(A_{2}\right)\right) \\
  - \left(P_{1} \left(A_{1}\right) + P_{1} \left(A_{1 2}\right)\right) \frac{P_{1} \left(A_{2}\right)}{1 - A_{2}} - P_{1} \left(A_{2}\right)\biggr].
  \label{eq:P2}
\end{multline}
$A_{1}$ is the same as above. The three additional arguments are $A_{1 2} = G^{2} \, V^{2} \left(\vec{R}_{1} - \vec{R}_{2}\right)$, $A_{2} = G^2 \, V^{2} \left(\vec{R}_{2}\right)$, and $A_{1 2 3} = G^3 \, V \left(\vec{R}_{1}\right) \, V \left(\vec{R}_{1} - \vec{R}_{2}\right) \, V \left(\vec{R}_{2}\right)$. The first term in Eq.~\eqref{eq:P2} generates not only all irreducible three-center diagrams, but also some that do not contain any three-center loops and are thus reducible. These are already generated by $F_{1}$ and are subsequently subtracted in order to prevent them from being counted twice.

For vanishing $r_{\mathrm{b}}$ and $\Im G < 0$, the second order generator \eqref{eq:F2} is proportional to $G^{2}$, \cf{} Ref.~\onlinecite{Elyutin:1980oq}. A numerical evaluation of the proportionality constant yields $F_{2} \left(G\right) \simeq (- 1.22338 + 1.63759 \ii) G^2$. For $r_{\mathrm{b}} > 0$, the simple proportionality is lost. The calculations are much more cumbersome, since, to solve Eq.~\eqref{eq:locator_expansion:selfconsistent_low}, the integral over $D_{2}$ has to be computed numerically for each value of $G$. Fig.~\ref{fig:fLambda_rydberg_rb_n10000} (black solid line) shows the results. Let us first comment on the performance of the method at the spectrum's edges (\ie{} for large $|\lambda|$). For $0 \le r_{\mathrm{b}} \le 0.5$, the solution to the self-consistent low concentration expansion is almost unaffected by the inclusion of $F_{2}$. The reference, the first- and the second-order result lie virtually on top of each other. This is reasonable, since three-center-cluster eigenstates do rarely populate the spectrum's edges. For $r_{\mathrm{b}} = 0.75$, the solution deviates from the reference considerably. The result is plotted only for $\lambda < 5$ since the numerical accuracy of the $D_{2}$-integral decreases for larger $\lambda$. However, already in the numerically tractable regime $\lambda < 5$, the agreement is even lower than what is achievable within first-order approximation. The reasons for that are presently unclear.

Concerning the spectrum's center, we see that the inclusion of three-center correlations yields a skewed distribution. The second order solution is asymmetric with the maximum shifted correctly to the right; compared to the reference, though, the asymmetry is a little too pronounced. Better agreement in this region can presumably be achieved by including higher-order generating functions, $F_{l}$ with $l > 2$. An example for $l = 3$ is shown in Fig.~\ref{fig:paths}(d). However, already deriving the kernels of $F_{l}$---let alone the subsequent $3l$-fold numerical integration---is hard since the number of correction terms needed to eliminate unwanted reducible diagrams grows disproportionately with $l$.

\subsubsection{High Concentration Limit}
\label{sec:spectrum:locator_expansion:high_concentration}

We therefore consider an alternative method which effectively sums up a certain subclass of diagrams to arbitrary order in $l$. The low concentration approximation considers all paths that look like trees globally and like repeated loops between a few centers (two for $l = 1$ and up to three for $l \le 2$) locally. By contrast, the paths of the high concentration approximation also look like loops on a global scale, see Fig.~\ref{fig:paths}(c). This is because, in a first step, only those terms of Eq.~\eqref{eq:locator_expansion} are considered for which $k = l$, \ie{}
\begin{align}
  z \, \overline{G_{0 0} \left(z\right)} \simeq 1 + \sum_{l = 1}^{\infty} \frac{\rho^l}{z^{l + 1}} \idotsint_{D_{l}} H_{0 1} \dotsm H_{l 0} \; \mathrm{d} \vec{R}_{1} \dotsm \mathrm{d} \vec{R}_{l}
  \label{eq:locator_expansion:high}
\end{align}
with $\rho \equiv 1$. At this point, $\overline{G_{0 0} \left(z\right)}$ is only built up from journeys that are perfect (\ie{} non-recurring and non-self-intersecting) loops of the form $0 \to 1 \to 2 \to \cdots \to l \to 0$ that. In a second step, a better approximation in form of a self-consistent equation for $\overline{G_{0 0} \left(z\right)}$ is obtained by allowing for arbitrarily many perfect-loop sub-journeys, $z \, \overline{G_{l l} \left(z\right)}$, at every intermediate atom $l$ on the path and at the home atom:
\begin{multline}
  z \, \overline{G_{0 0} \left(z\right)} \simeq 1 + \sum_{l = 1}^{\infty} \frac{1}{z^{l + 1}} \idotsint_{D_{l}} H_{0 1} z \, \overline{G_{1 1} \left(z\right)} \\
    \dotsm H_{l 0} z \, \overline{G_{0 0} \left(z\right)} \; \mathrm{d} \vec{R}_{1} \dotsm \mathrm{d} \vec{R}_{l}.
  \label{eq:locator_expfansion:selfconsistent_high}
\end{multline}
Invoking $H_{i j} = V \left(\vec{R}_{i} - \vec{R}_{j}\right)$ and $\overline{G_{l l} \left(z\right)} = \overline{G_{0 0} \left(z\right)}$ one finds
\begin{multline}
  z \, \overline{G_{0 0} \left(z\right)} \simeq 1 + \sum_{l = 1}^{\infty} \overline{G_{0 0} \left(z\right)}^{\, \left(l + 1\right)} \idotsint_{D'_{l}} W \left(\vec{R}_{1}\right) \\
    \times W \left(\vec{R}_{2} - \vec{R}_{1}\right) \dotsm W \left(- \vec{R}_{l}\right) \; \mathrm{d} \vec{R}_{1} \dotsm \mathrm{d} \vec{R}_{l},
\end{multline}
where $W\left(\vec{R}\right) = \Theta\left(R - r_{\mathrm{b}}\right) V\left(\vec{R}\right)$ with Heaviside's step function $\Theta$ is the potential $V$ restricted to distances larger than $r_{\mathrm{b}}$. Correspondingly, the domain of integration can be enlarged to $D'_{l} = \bigl\{\left(\vec{R}_{1}, \dotsc, \vec{R}_{l}\right) \in \mathbb{R}^{3 l} \, \big| \, R_{i} > r_{\mathrm{b}}, 1 < i < l, \, $ $\mathrm{and}$ $ \, \abs*{\vec{R}_{i} - \vec{R}_{j}} > r_{\mathrm{b}}, 1 < j + 1 < i \le l\bigr\}$. Because we were unable to make use of the upper equation in its present form, we approximated it further by letting the integration extend over the entirety of $\mathbb{R}^{3 l}$ \footnote{In other circumstances \cite{Martin-Mayor:2001kx}, this approximation has been referred to as the ``superposition approximation''.}. This is exact for $l \le 2$, but becomes somewhat ambiguous for larger $l$---ultimately, the step is justified by the results. The integral can then be rewritten as $W^{*\left(l + 1\right)} \left(\vec{0}\right)$, the $\left(l+1\right)$-fold convolution of the restricted potential $W$ with itself evaluated at the origin $\vec{0}$. By means of the convolution theorem one obtains
\begin{align}
  z \, \overline{G_{0 0} \left(z\right)} & \simeq 1 + \frac{1}{\left(2 \pi\right)^{3}} \int \frac{\overline{G_{0 0} \left(z\right)}^{\, 2} \left(\mathcal{F} \left\{W\right\}\right)^2 \left(\vec{K}\right)}{1 - \overline{G_{0 0} \left(z\right)} \, \mathcal{F} \left\{W\right\} \left(\vec{K}\right)} \, \mathrm{d} \vec{K}
  \label{eq:highcon}
\end{align}
where $\mathcal{F} \left\{W\right\}$ denotes the Fourier transform of $W$:
\begin{align}
  \mathcal{F} \left\{W\right\} \left(\vec{K}\right) & = \frac{9 \sqrt{3}}{2} \left(\frac{1}{3} - \frac{K_{Z}^{2}}{K^{2}}\right) \frac{3 j_{1} \left(K r_{\mathrm{b}}\right)}{K r_{\mathrm{b}}}
  \label{eq:W_FT}
\end{align}
with $j_{1}$ the second spherical Bessel function of the first kind. The right hand side of Eq.~\eqref{eq:highcon} is only defined for $r_{\mathrm{b}} > 0$, so that the $\vec{K}$-integral converges. Then we have
\begin{widetext}
  \begin{align}
    z - \frac{1}{\overline{G_{0 0} \left(z\right)}} & \simeq - \frac{9 \sqrt{3}}{4 \pi^2 r_{\mathrm{b}}} \int_{0}^{1} \left(3 u^{2} - 1\right) \int_{0}^{\infty} K \left\{\left[\frac{9 \sqrt{3}}{2 K r_{\mathrm{b}}} \left(3 u^{2} - 1\right) \overline{G_{0 0} \left(z\right)} + \frac{1}{j_{1} \left(K r_{\mathrm{b}}\right)}\right]^{-1} \negthickspace - j_{1} \left(K r_{\mathrm{b}}\right)\right\} \mathrm{d} K \, \mathrm{d} u
    \label{eq:highcon_simplified}
  \end{align}
\end{widetext}
For non-vanishing $r_{\mathrm{b}}$, the equation can be solved numerically for $\overline{G_{0 0} \left(z\right)}$ with $z = \lambda + \ii \varepsilon$, $\varepsilon$ small, but positive. We again used the CMA-ES for this task.

The low concentration approximation can be used to derive the Wigner semicircle distribution from Gaussian orthogonal statistics \cite{Flambaum:2000ti}. As shown in Sec.~\ref{sec:spectrum:applicability_RMT:GOE}, the semicircle law resembles the numerically exact reference density for large $r_{\mathrm{b}}$. We therefore expect similar behavior for the solution of Eq.~\eqref{eq:highcon_simplified}. Indeed, for $r_{\mathrm{b}} = 0.25$, it follows the Wigner semicircle smoothly, \cf{} Fig.~\ref{fig:fLambda_rydberg_rb_n10000} (dashed line), which makes the high concentration approximation evidently less satisfactory than the low concentration approximation. Better results are achieved for $r_{\mathrm{b}} = 0.5$ and $0.75$: the spectral width is reproduced reasonably well, and the curves have the characteristic asymmetry, since correlations are taken into account. Yet, in comparison to the low concentration approximation, the skewness is less pronounced, particularly around the spectrum's center. For $r_{\mathrm{b}} = 0.75$, the most probable eigenenergy is almost zero in the high-concentration approximation, whereas the exact numerical value is distinctively positive. It is unclear which step in the derivation of Eq.~\eqref{eq:highcon_simplified} (\ie{} either the neglect of non-loop-like diagrams or the extension of the integration volume to the whole $\mathbb{R}^{3 l}$) contributes most to these differences. The two approximations may at least partially counterbalance each other: for instance, the diagram depicted in Fig.~\ref{fig:paths}(d) is reproduced by a loop diagram with $l = 5$ where the positions of the atoms $1$ and $4$ and of the atoms $2$ and $5$ are close to each other, in particular, closer than $r_{\mathrm{b}}$, Fig.~\protect\ref{fig:paths}(e).

In total, our results show that the numerically exact reference lies between the low and the high concentration approximations. Therefore, it must be assumed that aspects of both types of diagrams---recurring short and non-recurring long loops---play a role. An interpolation between the methods (albeit in a non-self-consistent fashion) is treated in Ref.~\onlinecite{Mezard:1999cv} for an unrelated class of Euclidean random matrices and shall not be discussed here.

\section{Conclusions}
\label{sec:conclusions}

We have presented a model of coherent dipolar energy transfer between resonant levels of ultra-cold Rydberg atoms, specifically, of the non-radiative exchange of an $\mathrm{P}$ excitation among a large number of randomly distributed atoms in $\mathrm{S}$ Rydberg states. For this article, our attention was devoted to the spectral structure of the disordered many-body Hamiltonian. We conducted a numerical survey on the eigenvalue statistics of clouds with a large number of atoms $N$, compared it to results of established random matrix theories, and ultimately supplemented it with various analytical treatments for the asymptotic limit $N \to \infty$. The analytical approaches were based on the Matsubara-Toyozawa locator expansion, a self-consistent, diagrammatic perturbation theory for the ensemble-averaged resolvent. The results of our study are relevant equally for the fields of Rydberg physics as well as theoretical statistical physics.

The distribution of couplings between unblockaded Rydberg atoms has an algebraically decaying tail with diverging second moment. We found that this leads to a number of interesting effects. Significantly, although close atomic proximity is rare, it is yet a statistically important event. A diverging coupling strength leads to pair localization, a phenomenon that hinders excitations strongly to visit or leave a pair of closely separated atoms. Recognizing pair localization is instrumental in understanding excitonic energy transport in frozen Rydberg clouds. Signatures of the effect are visible in the spectral statistics: The spectral density is decaying algebraically and the nearest-neighbor level spacings obey Poissonian statistics in the wings of the spectrum. Towards the spectrum's center, the spacings undergo a transition to universal Wigner-Dyson statistics, indicating the presence of a mobility edge, \ie{} a crossover to a region of dominance of delocalized eigenstates.

Interestingly, the dipole blockade effect can be leveraged to tune the effective size of the Rydberg atoms. This opens up the possibility to control the short-range order in the cloud over a wide range, which is compelling for at least two reasons: First, it is well-known that spectral and transport properties can be very different in ordered and disordered systems. Rydberg gases are hence ideal testbeds for theories of transport in disordered systems with dipole-dipole interactions. Second, a strong dipole blockade inhibits short distances. The degree of pair localization can thus be gradually reduced and its effect on transport be isolated.

In the strong blockade regime, spectral statistics are reminiscent of those of the universal Gaussian random matrix ensemble, whereas, for weak blockade, there is significant agreement with the statistical properties of the $1$-stable random matrix ensemble. This is remarkable given that both ensembles regard matrix elements as uncorrelated, whereas the elements of the Euclidean many-body Rydberg Hamiltonian of our model are correlated. The comparison therefore allowed us to demonstrate that correlations lead both to asymmetry of the spectral density and to a considerable shift of the mobility edges to lower absolute energies. Importantly, the latter indicates clearly that the Euclidean correlations have a localizing effect on transport.

We also focused on self-consistent perturbative methods and showed that with these it is possible to reproduce characteristic features of the spectral density, not only for vanishing dipole blockade, but also in the strong blockade regime. In particular, we were able to describe the effects of pair localization at the spectrum's edges and the skewness introduced by Euclidean correlations at the center of the spectrum. We discussed the compatibility of different diagrammatic approaches and calculated solutions to both the low and the high concentration approximation to the locator expansion. We found that both approximations have their respective range of validity: the low concentration approximation performs especially well in the weak blockade regime, whereas the high concentration approximation delivers promising results only for strong blockade. Both methods proved to be valuable for addressing spectral problems involving ensembles of large disordered Hamiltonians with long-range dipolar interactions.

In future work, the statistical properties of the excitation migration and the eigenstates (associated to the eigenvalues analyzed in this article) remain to be investigated. This will allow us to characterize how exactly excitation energy transport in ultra-cold Rydberg gases is suppressed by effects like pair localization discussed in this article. The results presented here suggest that the suppression of transport is inversely correlated with the strength of the Rydberg blockade. Qualitatively, we expect a smooth transition between sub-diffusive and diffuse transport upon varying of the Rydberg blockade radius. Steps to substantiate this claim involve the calculation of explicit transport quantities, for instance, the ensemble-averaged square displacement as a function of time, from which one can extract the type and the coefficient of the diffusion process.

%\section{Acknowledgements}
%\label{sec:acknowledgements}

\appendix

\section{Validity of The Two-Level Approximation}
\label{app:validity_2lvl_approx}

In this appendix, we show that
the energy shift from a DC electric field
prepares two isolated Rydberg levels
well separated from all other resonances
\cite{Singer:2004gg,Singer:2005hc}.
We find that the reduction to two states
per atom can be maintained even for most
Rydberg atom pairs that are accidentally close together.

Like in the main text,
we are interested in the case
where the Rydberg atoms are either
excited to a $\mathrm{P}$ state
or relaxed to an $\mathrm{S}$ state,
each state with equal principal quantum number $n$.
Neglecting hyperfine structure
(which is insignificant
for high-lying Rydberg states
\cite{Li:2003qf} and
vanishes for elements with zero nuclear spin
like ${}^{78}\mathrm{Rb}$),
the $\mathrm{P}$ manifold contains
six and the $\mathrm{S}$ manifold two states,
\nP[@][@][@] and \nS[@][@][@],
respectively, where
$\Jg = 1/2$, $\Je = 1/2, 3/2$
are total angular momentum quantum numbers
and $\mg = \pm 1/2$,
$\me = \pm 1/2, \pm 3/2$ ($\abs{\me} \le \Je$)
the corresponding $\uvec{Z}$ components.
In order to arrive at a two-level system,
we have to energetically separate these states.
Due to their small angular momenta $L$
and the resultant large quantum defects,
the degeneracy of some of the states is already removed.
For Rubidium-$85$, for instance,
the split in energy between 
\nS[46][1/2][] and either the \nP[46][1/2][]
or the \nP[46][3/2][] states is
$1.305$ or $1.341 \mathrm{cm}^{-1}$, respectively
\cite{Li:2003qf}.
In addition,
we can use the Rydberg atoms' extreme sensitivity
to electric fields as an implement to break
the degeneracy of the different magnetic components
of the \nP[46][3/2][] state.
The DC Stark effect introduces an energy shift
that, to lowest order,
is quadratic in the applied field strength $F$
\cite{Zimmerman:1979ul,Haseyama:2003yq}.

The field strength should be weak enough
to avoid mixing of adjacent states
into the \nS[46][][] and \nP[46][][] level manifolds,
but strong enough to ensure that the induced shifts are  
orders of magnitude larger
than the typical dipole-dipole interaction energies
between the atoms.
For ${}^{85}\mathrm{Rb}$ and $n = 46$,
we find that
$F = 2.5 \mathrm{V} \mathrm{cm}^{-1}$
is appropriate, see below.
For this field strength
(with a field vector pointing
in the direction of the quantization axis),
the resonance peak of the transition
between \nS[46][1/2][\abs{1/2}] and either
\nP[46][1/2][\abs{1/2}], \nP[46][3/2][\abs{1/2}],
or \nP[46][3/2][\abs{3/2}]
is shifted to approximately
$1.291$, $1.324$, or $1.327 \mathrm{cm}^{-1}$,
respectively.
There are no avoided-crossing points
with adjacent level manifolds
in the vicinity of these states
\cite{Haseyama:2003yq}.

In the following, we show that,
to good approximation,
the subspace spanned by
\nS[46][1/2][1/2] and \nP[46][3/2][3/2]
(or, alternatively, \nS[46][1/2][-1/2]
and \nP[46][3/2][-3/2]),
is closed under Hamiltonian evolution
and thus that a description
reduced to these two levels alone
provides already an accurate account
of the excitation exchange processes in the gas.
In the presence of a single \nP[46][3/2][3/2] excitation
(among $(N-1)$ \nS[46][1/2][1/2] excitations),
the truncated state space has a dimension of $N$,
where $N$ is the number of Rydberg atoms in the cloud.
In this example,
we assume $N = 10^{4}$ as in the main text.
The density of the cloud
will be $\rho = 2.5 \times 10^{7} \mathrm{cm}^{3}$,
corresponding to an excitation volume
measuring $0.4 \mathrm{mm}^{3}$.
Due to the small size of the cloud
(and the magnitude of the involved resonance frequencies),
the interaction
between two Rydberg atoms $i$ and $j$
is dominated by dipole-dipole forces
proportional to their inverse cubed distance,
\begin{multline}
  \prescript{}{i j}{\bracket{\mathrm{P} \mathrm{S}}{\Htwo}{\mathrm{S} \mathrm{P}}}_{i j} \\ = \frac{\mueg^2}{4 \pi \epsilon_{0}} \prescript{}{i j}{\bracket[\big]{\mathrm{P} \mathrm{S}}{\deg \cdot \bigl(\oten - 3 \, \uvec{R}_{i j} \circ \uvec{R}_{i j}\bigr) \cdot \dge[j]}{\mathrm{S} \mathrm{P}}}^{}_{i j} \, R_{i j}^{-3} \\ = \beta \mathcal{A}\bigl(\uvec{R}_{i j}\bigr) \, \mathcal{R}\left(R_{i j}\right)
  \label{eq:exexchange}
\end{multline}
with $\epsilon_{0}$ the vacuum permittivity,
\deg{}, \dge{} the irreducible dipole transition operators
of rank one \cite{Wigner:1959vn,*Eckart:1930kx,*Edmonds:1996kq},
$\beta = \mueg^2 / \bigl(36 \sqrt{3} \epsilon_{0}\bigr) \simeq 2.59 \times 10^{-14} \mathrm{cm}^2$
(for ${}^{85}\mathrm{Rb}$, $n = 46$,
and $\mueg \simeq 2.389 \times 46^{2} \mathrm{D}$
the reduced dipole matrix element),
and the abbreviations
$\mathrm{S} = \nS[46][1/2][1/2]$ and
$\mathrm{P} = \nP[46][3/2][3/2]$.
$\mathcal{A}$ and $\mathcal{R}$
are defined in
Sec.~\ref{sec:theoretical_description:hamiltonian}
in the main text.
Terms proportional to $R_{i j}^{-2}$ and $R_{i j}^{-1}$
can be neglected.

\begin{figure}
  \begingroup%
  %\tikzexternalenable%
  (a) %\input{figures/2lvlvalidity_pair_a.tex}%
  \raisebox{-14.40813pt}{\includegraphics{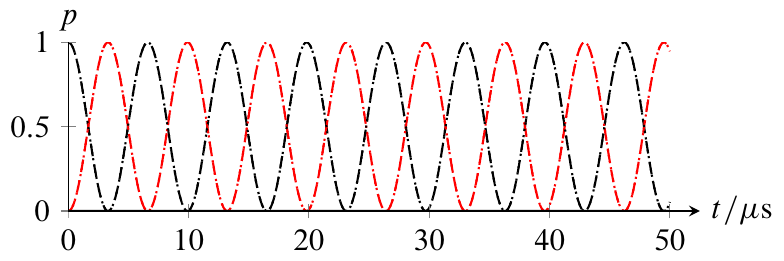}} \\%
  (b) %\input{figures/2lvlvalidity_pair_b.tex}%
  \raisebox{-14.40813pt}{\includegraphics{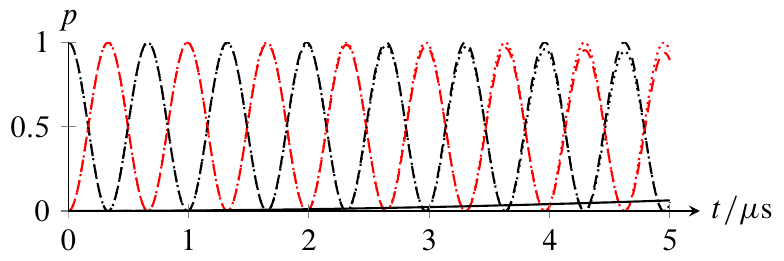}} \\%
  (c) %\input{figures/2lvlvalidity_pair_c.tex}%
  \raisebox{-14.40813pt}{\includegraphics{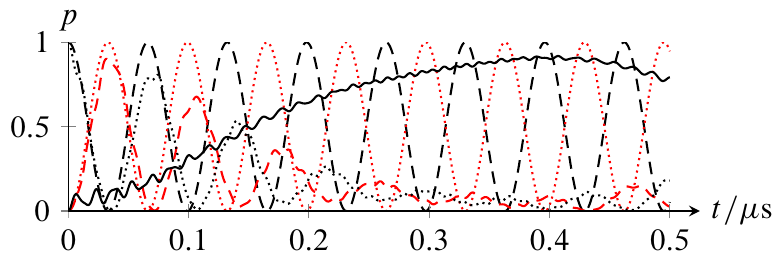}}%
  \endgroup%
  \caption{(color online) Breakdown of the
    two-level approximation for small distances.
    Shown are the coherent population dynamics
    of a pair of Rubidium-$85$ Rydberg atoms
    at different interatomic distances $R$,
    (a) $18.5 \mu\mathrm{m}$,
    (b) $8.60 \mu\mathrm{m}$, and
    (c) $3.99 \mu\mathrm{m}$,
    but for equal orientation
    $\protect\uvec{R} = (-1, 0, \sqrt{2})^{t} / \sqrt{3}$.
    In each case, the system is initialized in the state
    $\ket{\mathrm{P}, \mathrm{S}} = \ket*{\nP[46][3/2][3/2], \nS[46][1/2][1/2]}$.
    Plotted are the populations
    $p_{\mathrm{P} \mathrm{S}}$ and $p_{\mathrm{S} \mathrm{P}}$
    of the states $\ket{\mathrm{P}, \mathrm{S}}$ and
    $\ket{\mathrm{S}, \mathrm{P}}$, respectively,
    obtained in two-level approximation
    (black dashed and red/gray dotted)
    and when treated with the full Hamiltonian
    (black dotted and red/gray dashed).
    For the sake of error estimation, we also include
    $1 - p_{\mathrm{P} \mathrm{S}} - p_{\mathrm{S} \mathrm{P}}$
    (black solid).
    The results are discussed in the text.
  }%
  \label{fig:2lvlvalidity:pair}%
\end{figure}

For the chosen density,
the typical (most probable) distance
between nearest neighbor atoms is
$\left(2 \pi \rho\right)^{-1/3} \simeq 18.5 \mu\mathrm{m}$.
For comparison,
the extent of the Rydberg wave function
is $46^{2} a_{0} \simeq 0.098 \mu\mathrm{m}$.
In the case
$\mathcal{A}\bigl(\uvec{R}_{i j}\bigr) = 9 \sqrt{3} / (8 \pi) \simeq 0.62$,
the interaction strength between Rydberg atoms
at that distance is about
$0.62 \times 2 \pi \beta \rho \simeq 2.53 \times 10^{-6} \mathrm{cm}^{-1}$,
\ie{} roughly $1/1000$ the energy splitting
between \nP[46][3/2][\abs{1/2}]
and \nP[46][3/2][\abs{3/2}].
Fig.~\ref{fig:2lvlvalidity:pair}(a)
compares the time evolution
in the two-level approximation
with that generated by the full Hamiltonian
(taking into account the six states in the \nP[46][][]
and the two states in the \nP[46][][] manifold).
We see there that,
for a pair of atoms separated by
the typical inter-atomic distance,
the two-level treatment is valid,
since the curves agree perfectly.
The two-level approximation
should also work for more atoms,
as long as all atom pairs have mutual distances
close to (or larger than) the typical distance.
We explicitly verified this for three atoms.

The two-level treatment fails
once the coupling strength becomes comparable
in magnitude to the energy splitting
between the magnetic components
of the \nP[46][][] manifold.
This can happen
for rare, isolated pairs of atoms 
that are separated by much smaller distances
than the typical one.
On average, $66$ (out of $10^{4}$) atoms
have their nearest neighbor less than
$\left(100 \times 2 \pi \rho\right)^{-1/3} \simeq 3.99 \mu\mathrm{m}$
away.
For these short distances,
the dipole-dipole coupling
can become larger than
the DC Stark splitting.
The quantitative failure of
the two-level treatment
is illustrated in Fig.~\ref{fig:2lvlvalidity:pair}(c).
We see there that,
at this close distance,
the population is rapidly
transferred outside the reduced manifold.
This is different in Fig.~\ref{fig:2lvlvalidity:pair}(b),
where, compared to the Rabi period,
the evacuation of the reduced manifold
is a much slower process.

An important qualitative feature
of our two-level theory is 
the transition between delocalized
and pair-localized eigenstates
under variation of the eigenenergy,
see Sec.~\ref{sec:spectrum:mobility_edge}.
According to Fig.~\ref{fig:fS_rydberg_transition},
virtually all eigenstates with energies
$\abs{\Lambda} > 100 \beta \rho \simeq 6.48 \times 10^{-5} \mathrm{cm}^{-1}$
(and a large fraction of states
with $\abs{\Lambda} > 10 \beta \rho$)
should be localized.
Assuming $\mathcal{A}\bigl(\uvec{R}_{i j}\bigr) \simeq 0.62$,
this energy corresponds to an inter-atomic distance
of $6.28 \mu\mathrm{m}$,
which is right in between the situations
depicted in Figs.~\ref{fig:2lvlvalidity:pair}(b)
and (c).
That suggests that the energy regime
of the transition to pair-localized states
is fully covered by the simple two-level model.
For energies higher in absolute value, however,
the two-level approximation breaks down.
Notwithstanding,
it stands to reason that---%
although the eigenstates will not
be symmetric and antisymmetric superpositions
of $\ket{\nS[][][] \nP[][][]}$
and $\ket{\nP[][][] \nS[][][]}$---%
an \nP[][][] excitation will still be localized
at atom pairs as closely separated as 
$6.28 \mu\mathrm{m}$ (or less).
We see conclusive evidence for that
in numerical examples with
trimers of Rydberg atoms
propagated with the full Hamiltonian.

We thus come to the conclusion that
the two-level treatment works
for all but the closest atomic pairs.
Significantly, for the right choice of parameters
(chemical element, isotope, principal quantum number,
electric field strength, volume number density),
the two-level model is valid
at the most interesting energy scale,
where the system undergoes the transition
from delocalized to pair-localized states,
\cf{} Sec.~\ref{sec:spectrum:mobility_edge}.

\section{Derivation of the Probability Density $\boldsymbol{f_{H_{i j}}}$}
\label{app:probability_distributions}

For fixed $i$ and $j$, the probability density $f_{H_{i j}}$ of the off-diagonal matrix elements of the Hamiltonian \eqref{eq:H1} is identical to the density $f_{V}$ of the product $V = X Y$ of the independent random variables $X = \mathcal{A} \bigl(\uvec{R}_{i j}\bigr)$ (see Eq.~\eqref{eq:H1_Angular}) and $Y = \mathcal{R} \left(R_{i j}\right)$ (Eq.~\eqref{eq:H1_Radial}). The projection of $\uvec{R}_{i j}$ onto the $\uvec{Z}$-axis is uniformly distributed in the interval $[-1; 1]$. By changing variables one finds
\begin{align}
  f_{X} \left(x\right) & = \begin{dcases}
      \frac{1}{2 a \sqrt{\frac{1}{3} + \frac{x}{a}}}, & - \frac{a}{3} \le x \le \frac{2 a}{3} \\
      0, & \text{else}
    \end{dcases}
\end{align}
with the auxiliary constant $a = 27 \sqrt{3} / \left(8 \pi\right)$. The probability density function for the random variable $Y$ follows from the density \cite{Parry:2000fe,Tu:2002ga}
\begin{align}
  f_{R} \left(r\right) & = \begin{dcases}
      \frac{12 r^2 (d-r)^2 (2 d+r)}{\chi d^6}, & r_{\mathrm{b}} < r \le d \\
      0, & \text{else}
    \end{dcases}
  \label{eq:fR}
\end{align}
of finding a distance between $r$ and $r + \mathrm{d} r$ separating two atoms distributed uniformly inside a sphere of diameter $d = 2 \sqrt[3]{3 N / \left(4 \pi\right)}$ and outside spherical exclusion volumes with radius $r_{\mathrm{b}}$ around each atom. For the normalization $\chi$ one calculates
\begin{align}
  \chi & = 1 - \frac{r_{\mathrm{b}}^3 \left(8 d^3 - 9 d^2 r_{\mathrm{b}} + 2 r_{\mathrm{b}}^3\right)}{d^6}.
\end{align}
The density of $Y$ is therefore
\begin{align}
  f_{Y} \left(y\right) & = \begin{dcases}
      \frac{4 \left(d \sqrt[3]{y}-1\right)^2 \left(2 d \sqrt[3]{y}+1\right)}{d^6 \chi y^3}, & \frac{1}{d^3} \le y < \frac{1}{r_{\mathrm{b}}^3} \\
      0, & \text{else.}
    \end{dcases}
\end{align}
We obtain the product density $f_{V}$ by marginalizing out $Y$ from the joint probability density
\begin{align}
  f_{Y, V} \left(y, v\right) & = \frac{1}{y} \, f_{X} \left(\frac{v}{y}\right) f_{Y} \left(y\right)
\end{align}
of the variables $Y$ and $V = X Y$ \cite{Rohatgi:2011ap}. The result is defined piecewisely:
\begin{align}
  f_{V} \left(v\right) & = \begin{dcases}
      \int_{- 3 v / a}^{1 / r_{\mathrm{b}}^3} f_{Y, V} \left(y, v\right) \, \mathrm{d} y, & - \frac{a}{3 r_{\mathrm{b}}^3} < v \le - \frac{a}{3 d^3} \\
      \int_{1 / d^3}^{1 / r_{\mathrm{b}}^3} f_{Y, V} \left(y, v\right) \, \mathrm{d} y, & - \frac{a}{3 d^3} < v < \frac{2 a}{3 d^3} \\
      \int_{3 v / \left(2 a\right)}^{1 / r_{\mathrm{b}}^3} f_{Y, V} \left(y, v\right) \, \mathrm{d} y, & \frac{2 a}{3 d^3} \le v < \frac{2 a}{3 r_{\mathrm{b}}^3} \\
      0, & \text{else.}
    \end{dcases}
\end{align}
For $r_{\mathrm{b}} > 0$, straightforward integration yields
\begin{widetext}
  \begin{multline}
    f_{H_{i j}} \left(h\right) =
      \frac{64 N}{1485 b^3 \left(2 - b^2 \left(9 - 8 b + b^4\right)\right)} \\[\jot]
    \times \left\{\begin{aligned}
      & \begin{multlined}[b][10cm]
        u^{-3} \bigl[-216 \sqrt{\pi} b^2 \left(-u\right)^{2/3} \Gamma \left(\tfrac{4}{3}\right) \Gamma^{-1} \left(\tfrac{5}{6}\right) \\[\jot]
        + \sqrt{1 + u} \bigl(11 \left(8 + u \left(-4 + 3 u + 10 b^3 \left(-2 + u\right)\right)\right) \\[\jot]
        + 27 b^2 \left(16 + \left(8 - 5 u\right) u - 16 \, {}_{2} F_{1} \left(-\tfrac{1}{6}, 1; \tfrac{1}{3}; -u\right)\right)\bigr)\bigr],
      \end{multlined} && \; -1 < u \le -b^{-3} \\[\jot]
      & \begin{multlined}[b][10cm]
        u^{-3} \bigl[-8 \sqrt{1 + b^3 u} \left(65 + b^3 u \left(-6 + b^3 u\right)\right) \\[\jot]
        + \sqrt{1 + u} (11 \left(8 + u \left(-4 + 3 u + 10 b^3 \left(-2 + u\right)\right)\right) \\[\jot]
        + 27 b^2 \left(16 + \left(8 - 5 u\right) u\right)) \\[\jot]
        + 432 \left(-b^2 {}_{2} F_{1} \left(-\tfrac{2}{3}, \tfrac{1}{2}; \tfrac{1}{3}; -u\right)
          + {}_{2} F_{1} \left(-\tfrac{2}{3}, \tfrac{1}{2}; \tfrac{1}{3}; -b^3 u\right)\right)\bigr],
      \end{multlined} && \; -b^{-3} < u < 0 \lor 0 < u < 2 b^{-3} \\[\jot]
      & \begin{multlined}[b][10cm]
        \tfrac{55}{14} \left(-7 + b^2 \left(27 - 21 b + b^7\right)\right),
      \end{multlined} && \; u = 0 \\[\jot]
      & \begin{multlined}[b][10cm]
        u^{-3} \bigl[-132 \sqrt{3} + 54 \sqrt[3]{2} \sqrt{3} b^2 u^{2/3} \left(-3 + 4 \, {}_{2} F_{1} \left(-\tfrac{1}{6}, 1; \tfrac{1}{3}; -2\right)\right) \\[\jot]
        + \sqrt{1 + u} \bigl(11 \left(8 + u \left(-4 + 3 u + 10 b^3 \left(-2 + u\right)\right)\right) \\[\jot]
        + 27 b^2 \left(16 + \left(8 - 5 u\right) u - 16 \, {}_{2} F_{1} \left(-\tfrac{1}{6}, 1; \tfrac{1}{3}; -u\right)\right)\bigr)\bigr],
      \end{multlined} && \; 2 b^{-3} \le u < 2 \\[\jot]
      & \begin{multlined}[b][10cm]
        0,
      \end{multlined} && \; \text{else}
    \end{aligned}\right.
    \label{eq:fH1}
  \end{multline}
\end{widetext}
with the abbreviations $u = 3 r_{\mathrm{b}}^3 h / a$, $b = \sqrt[3]{6 N / \pi} / r_{\mathrm{b}}$, and the hyper-geometric function ${}_{2} F_{1}$. \cite{Abramowitz:1972mi} Plots of Eq.~\eqref{eq:fH1} for $r_{\mathrm{b}} = 0$ and $0.75$ can be found in Fig.~\ref{fig:probability_distributions}. Eq.~\eqref{eq:fH1} can be used to calculate the expected values of $(H_{i j}){}^k$. The mean vanishes,
\begin{subequations}
  \begin{align}
    \overline{H_{i j}} & = 0,
      \label{eq:H1_mean}
    \intertext{but the variance is finite and strongly dependent on $r_{\mathrm{b}}$,}
    \overline{(H_{i j}){}^2} & =
      \frac{27 b^{6} \left(5 + b^{2} \left(-9 + 4 b\right) + 6 \ln \left(b\right)\right)}{160 N^{2} \left(-2 + b^{2} \left(9 - 8 b + b^{4}\right)\right)},
      \label{eq:H1_variance}
  \end{align}
\end{subequations}
with an algebraic divergence, $r_{\mathrm{b}}^{-3}$, in the limit $r_{\mathrm{b}} \to 0$.

\section{Limitations of The Super-atom Picture}
\label{app:limitations_superatoms}

The excitation processes in an ultra-cold Rydberg gas may involve more than two atoms \cite{Ates:2007nt,*Ates:2007rp,Younge:2009gt,Gurian:2012cy}, in which case the excitation is coherently (but not necessarily evenly) spread among the participants:
\begin{subequations}
  \begin{align}
    \ket{\mathrm{S}}_{A} & = \ee^{\ii \vec{K} \cdot \vec{R}_{A}} \sum_{i \in I_{A}} \sqrt{w_{A i}} \, \ee^{\ii \vec{K} \cdot \vec{r}_{A i}} \ket{\mathrm{S}}_{i}
    \label{eq:blockade_example_A}
    \shortintertext{and}
    \ket{\mathrm{P}}_{B} & = \ee^{\ii \vec{K} \cdot \vec{R}_{B}} \sum_{j \in I_{B}} \sqrt{w_{B j}} \, \ee^{\ii \vec{K} \cdot \vec{r}_{B j}} \ket{\mathrm{P}}_{j}
    \label{eq:blockade_example_B}
  \end{align}
\end{subequations}
are exemplary states of two non-intersecting collections $A, B$ (the index sets fulfill $I_{A} \cap I_{B} = \{\}$) of atoms sharing a single $\mathrm{S}$ and $\mathrm{P}$ interaction, respectively. The collections---sometimes also referred to as ``super-atoms''---are typically arranged in loose spheres, the so-called blockade spheres \cite{Tong:2004no}, with centers $\vec{R}_{A}, \vec{R}_{B}$ and radii approximately equal to $r_{\mathrm{b}}$. In the above equations, $w_{A i}, w_{B j}$ are normalized weights, $\vec{r}_{A i}, \vec{r}_{B i}$ relative atomic positions, and $\vec{K}$ the wave vector of the excitation. To see that these collections behave like bloated single Rydberg atoms, let us calculate the exchange matrix element $\prescript{}{A B}{\bracket{\mathrm{P} \mathrm{S}}{\Htwo}{\mathrm{S} \mathrm{P}}}_{A B}$. It reduces to
\begin{align}
  \frac{\mueg^2}{16 \pi \epsilon_{0}} \sum_{i \in I_{A}} \sum_{j \in I_{B}} w_{A i} w_{B j} \, \vec{\nabla}_{-1} \vec{\nabla}_{1} \frac{1}{\abs*{\vec{R}_{A B} + \vec{r}_{A B i j}}},
  \label{eq:exexhangeAB}
\end{align}
where $\vec{\nabla}_{\alpha}$ is the $\alpha$-th covariant cyclic component of the gradient operator, acting here with respect to $\vec{R}_{A B} = \abs*{\vec{R}_{A} - \vec{R}_{B}}$, and $\vec{r}_{A B i j}$ is defined as $\vec{r}_{A i} - \vec{r}_{B j}$. One further calculates
\begin{multline}
  \vec{\nabla}_{-1} \vec{\nabla}_{1} \frac{1}{\abs*{\vec{R}_{A B} + \vec{r}_{A B i j}}} = \frac{1}{2} \sum_{L = 0}^{\infty} \sum_{M = -L}^{L} \left(-1\right)^{M} \\
    \times \sqrt{\left(L - M + 1\right) \left(L - M + 2\right) \left(L + M + 1\right) \left(L + M + 2\right)} \\
    \times \mathfrak{I}_{L + 2, -M} \left(\vec{R}_{A B}\right) \, \mathfrak{R}_{L M} \left(-\vec{r}_{A B i j}\right) \\
  = \frac{1}{2} \left(3 \cos^2 \Theta_{A B} - 1\right) \, R_{A B}^{-3} + \dotsb
  \label{eq:laplace_expansion}
\end{multline}
with $R_{A B} > r_{A B i j}$ and
\begin{subequations}
  \begin{align}
    \mathfrak{I}_{L M} \left(\vec{r}\right) & = \sqrt{\frac{4 \pi}{2 L + 1}} \, r^{L} \, Y_{L M} \left(\uvec{r}\right)
    \shortintertext{the irregular and}
    \mathfrak{R}_{L M} \left(\vec{r}\right) & = \sqrt{\frac{4 \pi}{2 L + 1}} \, r^{-\left(L + 1\right)} \, Y_{L M} \left(\uvec{r}\right)
  \end{align}
\end{subequations}
the regular solid harmonic \cite{Varsalovic:1989on}. When the first term in the Laplace expansion, Eq.~\eqref{eq:laplace_expansion}, is inserted into Eq.~\eqref{eq:exexhangeAB}, the result indeed has the form of the usual exchange interaction between two single Rydberg atoms at the positions $\vec{R}_{A}$ and $\vec{R}_{B}$. If we define the multi-pole moment
\begin{align}
  \left\langle \mathfrak{R}_{L M} \right\rangle_{A} & = \sum_{i \in I_{A}} w_{A i} \mathfrak{R}_{L M} \left(\vec{r}_{A i}\right),
  \label{eq:multipoleA}
\end{align}
and accordingly for collection B, we can write
\begin{multline}
  \sum_{i \in I_{A}} \sum_{i \in I_{B}} w_{A i} w_{B i} \mathfrak{R}_{L M} \left(-\vec{r}_{A B i j}\right) = \sum_{L' = 0}^{L} \left(-1\right)^{L - L'} \smashoperator[l]{\sum_{M' = -L'}^{L'}} \\
    \times \sqrt{\tbinom{L + M}{L' + M'}} \sqrt{\tbinom{L - M}{L' - M'}} \, \left\langle \mathfrak{R}_{L - L', M - M'} \right\rangle_{A} \, \left\langle \mathfrak{R}_{L' M'} \right\rangle_{B}.
\end{multline}
The multi-pole moments are very simple in case of perfectly spherical, homogeneously excited collections. For $w_{A i} = \abs*{I_{A}}^{-1} = w_{B j} = \abs*{I_{B}}^{-1}$, they become $\left\langle \mathfrak{R}_{L M} \right\rangle_{A} = \left\langle \mathfrak{R}_{L M} \right\rangle_{B} = 3 r_{\mathrm{b}}^L / \left(L + 3\right)$. A numerical analysis with uniformly sampled $\uvec{R}_{A B}$ (over a unit sphere's surface) showed that the next non-vanishing term in the Laplace expansion, $L = 2$, must be included, if $R_{A B}$ is smaller than roughly $2.5 r_{\mathrm{b}}$. Thus, for sufficiently large $R_{A B}$, the blockade spheres are indistinguishable from single Rydberg atoms. For very small $R_{A B}$, both, first the expansion and then the blockade sphere picture, break down. The latter, because Eqs.~\eqref{eq:blockade_example_A} and \eqref{eq:blockade_example_B} are only justifiable, if the collections are isolated and well separated. Should they intersect, then they may share the two Rydberg excitations in a non-trivial way.

%\bibliography{rydberg}

%

\end{document}